\documentclass{aa} 
\usepackage{graphicx} 
\usepackage{txfonts}
\usepackage{xcolor}
\usepackage{url} 
\usepackage{subcaption}
\usepackage{array}
\usepackage{threeparttable} 
\usepackage{threeparttablex} 
\usepackage{comment}
\usepackage{longtable}
\usepackage{twoopt}
\usepackage{natbib}
\bibpunct{(}{)}{;}{a}{}{,} 
\usepackage{hyperref}    
\makeatletter
  \newcommandtwoopt{\citeads}[3][][]{\href{http://adsabs.harvard.edu/abs/#3}%
    {\def\hyper@linkstart##1##2{}%
     \let\hyper@linkend\@empty\citealp[#1][#2]{#3}}}
  \newcommandtwoopt{\citepads}[3][][]{\href{http://adsabs.harvard.edu/abs/#3}%
    {\def\hyper@linkstart##1##2{}%
     \let\hyper@linkend\@empty\citep[#1][#2]{#3}}}
  \newcommandtwoopt{\citetads}[3][][]{\href{http://adsabs.harvard.edu/abs/#3}%
    {\def\hyper@linkstart##1##2{}%
     \let\hyper@linkend\@empty\citet[#1][#2]{#3}}}
  \newcommandtwoopt{\citeyearads}[3][][]%
    {\href{http://adsabs.harvard.edu/abs/#3}
    {\def\hyper@linkstart##1##2{}%
     \let\hyper@linkend\@empty\citeyear[#1][#2]{#3}}}
\makeatother

\begin{document}


\title{Activity and composition of periodic comets 67P/Churyumov-Gerasimenko and 103P/Hartley 2 at two different perihelion passages}

\author{E. Hemmen\inst{1},
        M. Vander Donckt\inst{1},
        E. Jehin\inst{1},
        S. Hmiddouch\inst{1,2},
        K. Aravind\inst{1},
        J. Manfroid\inst{1},
        Z. Benkhaldoun\inst{3,2},
        A. Jabiri\inst{2},
        \and
        S. Ganesh\inst{4}    
        }

\institute{Space Sciences, Technologies and Astrophysics Research (STAR) Institute, University of Liège, Liège, Belgium\\
        \email{elisehemm@outlook.com}
        \and
        Oukaimeden Observatory, High Energy Physics, Astrophysics and Geoscience Laboratory, Faculté des Sciences Semlalia (FSSM), Cadi Ayyad University, Marrakesh, Morocco.
        \and
        Department of Applied Physics and Astronomy, and Sharjah Academy for Astronomy, Space Sciences and Technology, University of Sharjah, United Arab Emirates.
        \and
        Astronomy and Astrophysics division, Physical Research Laboratory, Ahmedabad 380009, India}

\titlerunning{67P and 103P at two perihelion passages}
\authorrunning{E. Hemmen et al.}

\abstract
{Comets are among the most primitive Solar System bodies, and they preserve material from its early formation stages. Studying their activity and composition provides key insights into the physical and chemical conditions of the primordial solar nebula. Periodic comets, such as 67P/Churyumov–Gerasimenko and 103P/Hartley 2, offer the opportunity for comparative and long-term monitoring over successive perihelion passages.} 
{Through photometry and spectroscopy, we studied the evolution of the activity and chemical composition of comet 67P during its 2025 and 2021 perihelion passages and of comet 103P during its 2010 and 2023 passages. For each comet, we aim to compare their behavior from one apparition to another.}
{We used the TRAPPIST telescopes to monitor the comets using broadband and narrowband filters. From the broadband images, we produced light curves and computed color indices for each passage, and we derived the activity slopes. We used a Haser model to compute the production rates of five gaseous species (CN, C$_2$, C$_3$, OH, and NH) and derived the proxy parameter $Af\rho$ for dust activity. We also observed both comets in spectroscopy during their most recent apparition using the Himalayan Chandra Telescope and compared the spectroscopic data to our results obtained through photometry.}
{For both comets, our analysis of coma colors does not reveal any significant change from one passage to the other, indicating that the properties of the released dust grains are similar. Our values of the color indices are consistent with the mean values for Jupiter-family comets. We measured a slight increase in the gas and dust activities of comet 67P between 2015 and 2021, probably due to the small change in the comet's orbit that led the perihelion distance to decrease from 1.24 au for the first apparition to 1.21 au for the second one. Regarding 103P, we unambiguously measured a decrease (of at least 50\%) in the gas and dust activities between 2010 and 2023, showing a different behavior for this young, active comet. We find a typical chemical composition for both comets and detect no variation of the C$_2$-to-CN production rate ratios and dust-to-gas ratios from one passage to the other, indicating constant compositions, even if the level of activity has changed for 103P.}
{}


\maketitle

   
\section{Introduction}

Comets play a crucial role in exploring the Solar System's history, as they are considered to be among its most primitive bodies. Since their formation about 4.6 billion years ago, cometary nuclei have been preserved in the cold, remote regions of the Solar System, and thus their chemical composition has remained nearly unchanged. Mainly under the influence of gravitational forces, some of these nuclei are sometimes deflected toward the planetary region. As they approach the Sun, the ices they contain sublimate under the effect of solar heat, releasing grains of cometary dust and creating a large envelope around the nucleus, known as the coma. This coma might be large and bright, allowing the study of the comet nucleus composition from the Earth. The study of these small celestial bodies and their chemical composition is of great importance since it gives some clues about the solar nebula's material and offers a unique window on the conditions that were prevailing at the time of Solar System formation \citep{de2015planetary, swamy2010physics, 2020icpr.book.....T}. 

Our targets, 67P/Churyumov-Gerasimenko (hereafter 67P) and 103P/Hartley 2 (hereafter 103P), are periodic comets of the Jupiter family with orbital periods of 6.4 and 6.5 years, respectively. Both comets were observed extensively by two very successful missions, i.e., the ESA Rosetta mission\footnote{\url{https://www.esa.int/Science_Exploration/Space_Science/Rosetta}} and the NASA EPOXI mission\footnote{\url{https://science.nasa.gov/mission/deep-impact-epoxi/}}. They are thus prime targets of high interest for the community. They have both been observed during two distinct perihelion passages with TRAPPIST (TRAnsiting Planets and PlanetesImals Small Telescope). The latter consists of a pair of twin 60-cm telescopes: TRAPPIST-South (TS, I40), installed in 2010 at ESO's La Silla Observatory in Chile, and TRAPPIST-North (TN, Z53), installed in 2016 at the Oukaïmeden Observatory in Morocco. Using these telescopes, observations can be performed from both the southern and northern hemispheres of the Earth, and thus it is possible to access the whole sky at anytime. The TRAPPIST project is dedicated to the detection of transiting exoplanets and the study of Solar System planetesimals \citep{2011Msngr.145....2J}. It notably allows one to conduct long-term monitoring of comets along their orbits, before and after perihelion, and to obtain highly homogeneous datasets. This is essential to studying the evolution of cometary activity, as comets receive different amounts of sunlight along their orbit, and to ensuring a reliable comparison of the activity and chemical composition between different comets or different passages. 

In this work, we used TRAPPIST data to determine the gas production rates of comets 67P and 103P and estimate their dust activity during two different perihelion passages. We also used the 2-m Himalayan Chandra Telescope (HCT) to observe both comets in spectroscopy during their recent apparitions. The major aim was to spot similarities and differences in the behavior of the two comets from one passage to another using the same instrumentation and the same measurement and computation techniques for both apparitions. The observation methods and conditions are given in Section \ref{section:obs}. We describe our datasets as well as the whole process used for data reduction. We then present the results of our calculations on the gas production rates of various volatile species (OH, NH, C$_2$, C$_3$, and CN) and dust outgassing in Section \ref{section:results}, and discuss them in Section \ref{section:discu}.


\section{Observations and data reduction}
\label{section:obs}

The comets 67P and 103P were observed using two different facilities, TRAPPIST and HCT. While TRAPPIST observed the comets in photometry across multiple epochs during two different apparitions, HCT was used to observe the comets using long-slit spectroscopy during their recent apparitions. The following sub-sections describe the observation and reduction methods employed by both facilities.

\subsection{Photometry - TRAPPIST}

TRAPPIST telescopes are equipped with highly sensitive charge coupled device (CCD) cameras of 2048 $\times$ 2048 15-µm pixels covering a field of view of 22' $\times$ 22'. Comet observations are performed with filters that isolate narrow wavelength bands corresponding to the emission bands of OH, NH, CN, C$_3$, and C$_2$ (i.e., the main emission bands observed in the optical spectrum of comets) and to the emission-free regions BC, GC, and RC characterizing the sunlight reflected by dust grains (see \cite{2000Icar..147..180F} for the characteristics of the HB filters). In addition, cometary images are also taken with the Johnson-Cousin B, V, R, and I broadband filters. 

Comet 67P was observed with TS during its 2015 passage at the same time as the Rosetta space mission. The perihelion occurred on August 13 at a heliocentric distance ($r_h$) of 1.24 au. Despite the difficulty of observing the comet from any Earth-based site due to the low solar elongation at that apparition, 395 images could be collected during 79 observation nights spread between April 2015 ($r_h = 1.8$ au) and July 2016 ($r_h = 3.4$ au). However, most of these images were taken with broadband filters, while very few images were collected with narrowband filters. In 2021, the comet passage was monitored simultaneously by TS and TN. The perihelion occurred on November 2 at a heliocentric distance of 1.21 au, and the observations were performed during 67 nights from May 2021 ($r_h = 2.3$ au) to February 2022 ($r_h = 1.8$ au). In total, 681 images were collected. 

Comet 103P was observed with TS during its passage close to Earth in 2010. The perihelion occurred on October 28 at a heliocentric distance of 1.06 au, but the observations only started a few weeks later, as these were the first scientific data collected by TS, which had its first light a couple of months before. In consequence, no pre-perihelion data were available for this passage. However, a highly intensive data collection conducted with the telescope during 64 nights spread between December 2010 ($r_h = 1.2$ au) and May 2011 ($r_h = 2.5$ au) yielded a total of 1792 images, most of them taken with narrowband filters. Due to a bad position in the sky during its perihelion passage in 2017, the comet could not be observed from any place on Earth, but it was observed during its latest passage in 2023. The perihelion occurred on October 12 at a heliocentric distance of 1.06 au. The observations with both TS and TN were conducted during 68 nights between June 2023 ($r_h = 1.8$ au) and March 2024 ($r_h = 2.1$ au), obtaining 709 images of the comet. 

Data calibration was realized following the standard process, which is the same procedure as explained in previous TRAPPIST publications \citep{Hmiddouch25, VanderDonckt26}, using Python scripts combined with the Image Reduction and Analysis Facility (IRAF). The reduction of the narrowband filter images, including bias and dark subtraction and flat-field correction, was first performed using up-to-date master bias, dark, and flat frames, respectively. Then, the sky contribution (i.e., light flux mainly due to the Moon, twilight, atmospheric activity, and light pollution coming from the Earth's surface) had to be removed. This flux is generally difficult to estimate, especially in the case of extended objects such as comets, but the large ($22' \times 22'$) TRAPPIST's field of view makes this task easier. Indeed, we used the median flux computed at the closest nucleocentric distance from the coma optocenter, where there were no cometary emissions as the background flux, and then we removed it from the whole image. After the sky background subtraction, we obtained median radial profiles based on each comet image, and we used scaled BC profiles to correct the dust-contaminated flux measured with the gas filters. The scaling factors used depend on the importance of dust contamination in each gas filter (OH: 19; NH: 24; CN: 30; C$_2$: 170; C$_3$: 248). The choice of BC images for this correction is relevant since the flux in the blue continuum is not itself contaminated by gas emissions. Finally, the absolute flux calibration was performed based on observations of standard stars with the HB cometary filters, according to the formula developed by \cite{2000Icar..147..180F}. The reduction of the broadband filter images followed the same process. We used the La Palma extinction curve for TN and the La Silla curve for TS. We observed Stetson standard fields\footnote{\url{https://www.canfar.net/storage/list/STETSON/Standards}} regularly and computed the zero points for each filter.

In this whole process of data reduction and flux calibration, the most significant sources of uncertainty come from the sky contribution removal and absolute flux calibration. Consequently, for the sky contribution subtraction, we computed a 3$\sigma$ uncertainty on the estimated sky flux. For the absolute flux calibration, the uncertainty is due to the fact that the atmospheric extinction coefficients were determined from the typical extinction curve of the observation site (La Silla or Oukaïmeden), although in reality these values vary slightly from one night to another. We thus considered a 5\% uncertainty on the extinction coefficients, based on the 5\% scattering of the coefficients determined through long-term observations of standard stars. In particular, this uncertainty becomes significant at high airmass. In the following sections, the mentioned uncertainties correspond to a quadratic combination of the 3$\sigma$ uncertainty on the sky contribution estimation and the 5\% uncertainty on the extinction coefficients. 

\subsection{Long-slit spectroscopy - HCT}

Spectroscopic observations were performed with the Hanle Faint Object Spectrograph and Camera (HFOSC) mounted on the 2.0-m HCT at the Indian Astronomical Observatory (IAO; see \cite{2022Icar..38315042A}). A slit 11' long and 1.92" wide, providing a resolving power of approximately 1300 (corresponding to a spectral resolution of about 1.45 \AA~per pixel), was used for the comet observations. The observation of comet 67P was obtained on the night of November 9, 2021, when the comet was at heliocentric and geocentric distances of 1.21 au and 0.42 au, respectively. The observation of comet 103P was carried out on September 28, 2023, when the comet was at heliocentric and geocentric distances of 1.08 au and 0.38 au, respectively.

For each epoch, both comets and sky frames were recorded with exposure times of 1200 s each. The comets were positioned at the center of the slit, oriented along the east-west direction. The spectrophotometric standard star BD+28 4211 was observed together with comet 67P, and HD 74721 with comet 103P. Both standard stars were observed with a much wider slit (15.4") to minimize light losses and to derive the instrumental sensitivity function required for flux calibration. Halogen-lamp spectra, zero-exposure frames, and FeAr-lamp spectra were acquired for flat-field correction, bias subtraction, and wavelength calibration, respectively. During both observing runs, the solar analog HD 81809 (G2 V; \citep{2000Icar..147..180F}) was observed with the same instrumental setup as the comets to enable continuum removal from the cometary spectra.

\begin{figure}[h!]
    \centering
        \includegraphics[width=\columnwidth]{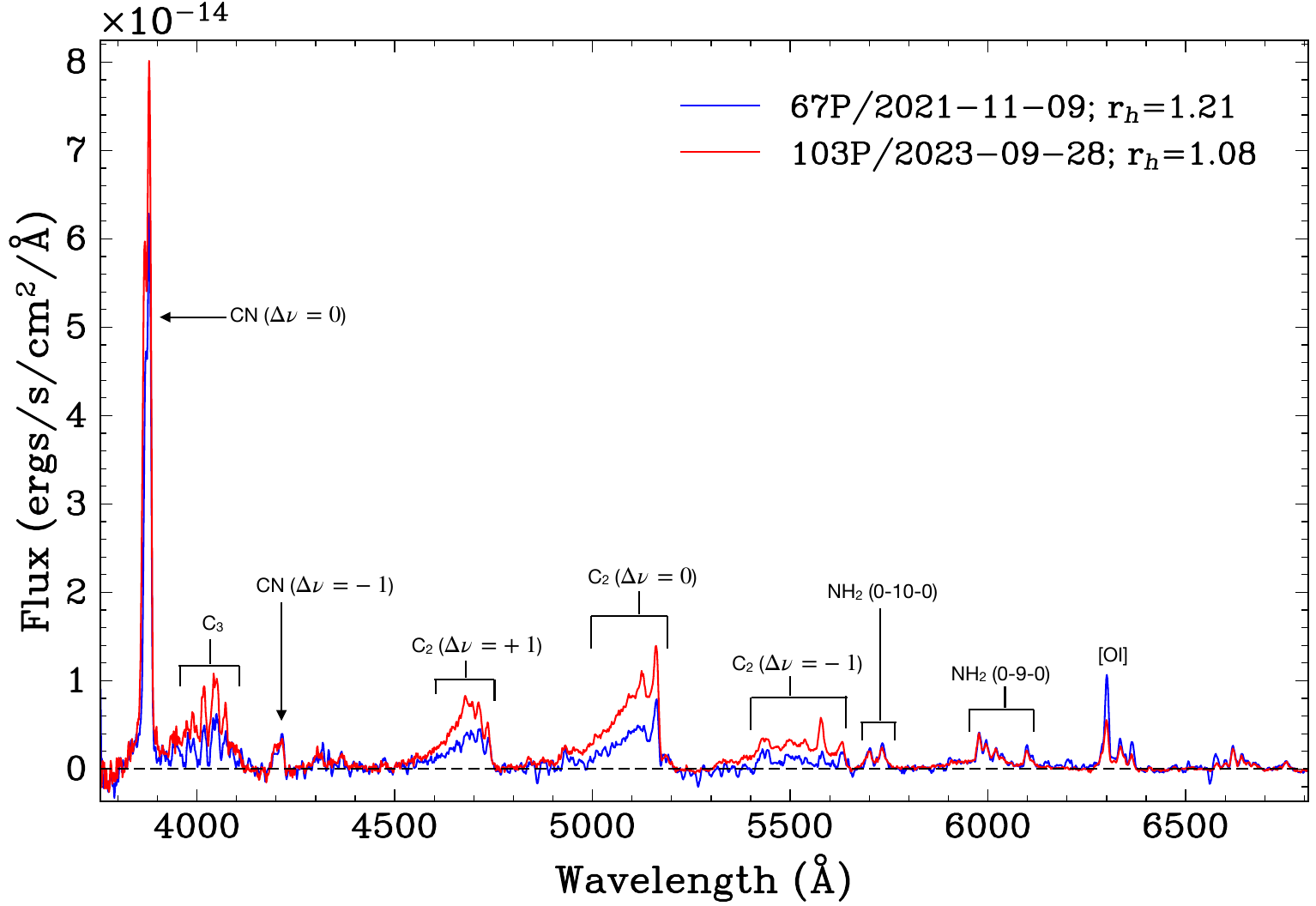} 
    \caption{Flux-calibrated, continuum-subtracted optical spectra of comets 67P and 103P observed from HCT on November 9, 2021, and September 28, 2023, respectively, extracted within a 60" aperture centered on the comet.}
    \label{spectra}
\end{figure}

The comet spectra were reduced and calibrated using custom Python scripts in combination with IRAF, following standard long-slit reduction procedures described in \citet{2022Icar..38315042A}. The spectra of the two comets were extracted from the reduced frames corresponding to their respective epochs using the Python routines. The standard-star spectra were used to determine the characteristic spectral trace of the instrument, which, with necessary adjustments, was applied to trace and extract the comet spectra along the dispersion axis. The standard-star spectra were extracted with the IRAF task "apall", which provides effective sky subtraction using regions on both sides of the target. Wavelength and flux calibrations were performed using standard IRAF packages. A suitably scaled and slope-corrected solar spectrum (see \cite{2022Icar..38315042A} for details) was applied to remove the reflected solar continuum contribution. The flux-calibrated, continuum-subtracted spectra of the two comets, extracted within a 60" aperture centered on the comet, are shown in Fig.~\ref{spectra}.

\section{Data analysis and results}
\label{section:results}

\subsection{Analysis of the light curves and coma dust colors}

In this section, we present the results of our magnitude and color analysis of comets 67P and 103P using images acquired with the B, V, R, and I broadband filters. The magnitude values given here were obtained by integrating the cometary flux inside a 5-arcsec circular aperture. 

\begin{figure}[htbp]
    \centering
        \includegraphics[width=\linewidth]{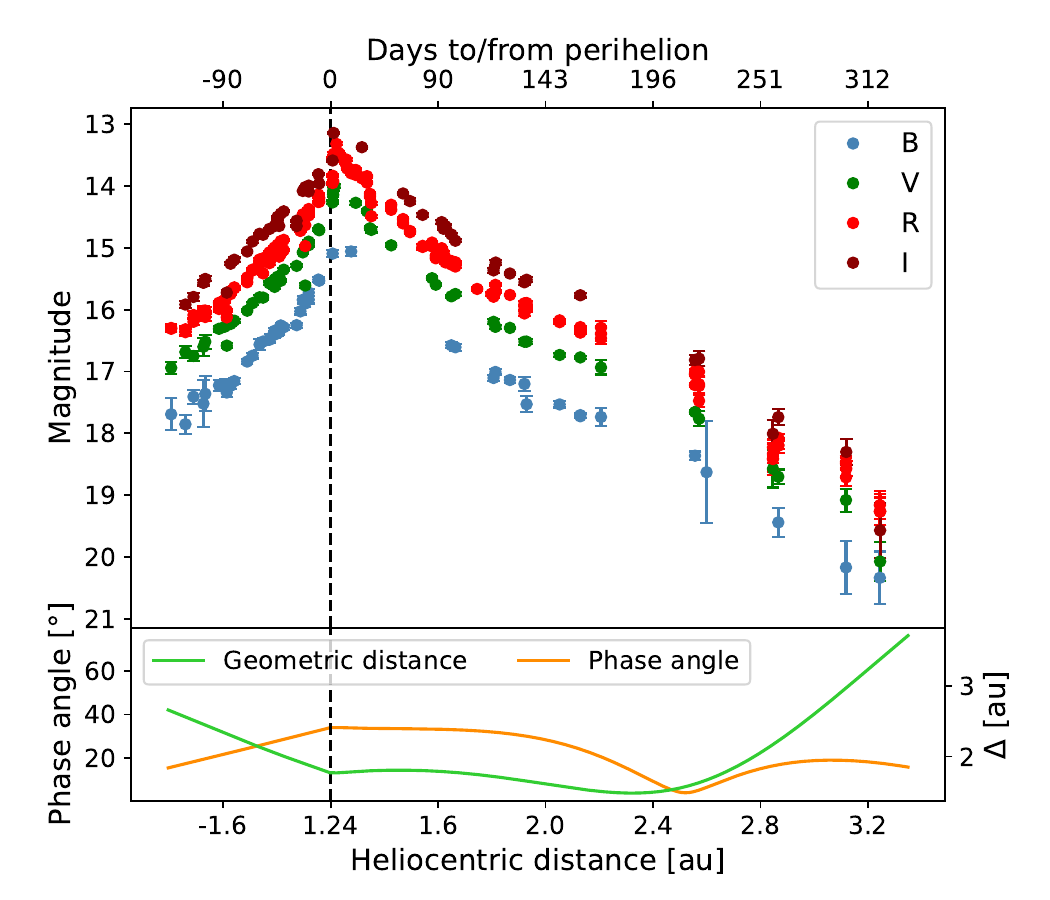} \\
        \includegraphics[width=\linewidth]{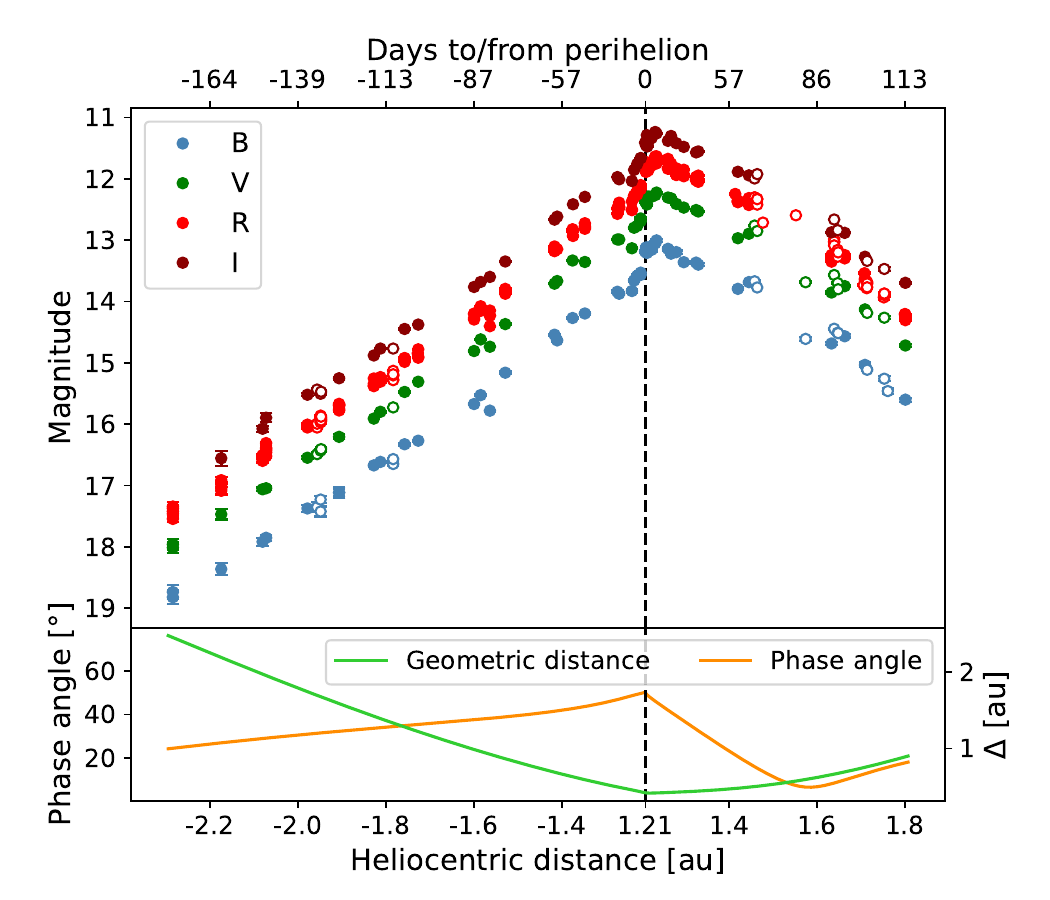}
    \caption{Light curves of comet 67P measured within a 5-arcsec aperture as a function of heliocentric distance and time during its 2015 (top) and 2021 (bottom) perihelion passages. The dotted black line indicates the perihelion at 1.24 au on August 13, 2015, and 1.21 au on November 2, 2021. Filled and open points stand for TS and TN data, respectively. The evolution of geometric distance and phase angle of the comet during the observations is shown.}
    \label{fig:mag_67P}
\end{figure}

\begin{figure}[htbp]
    \centering
        \includegraphics[width=\linewidth]{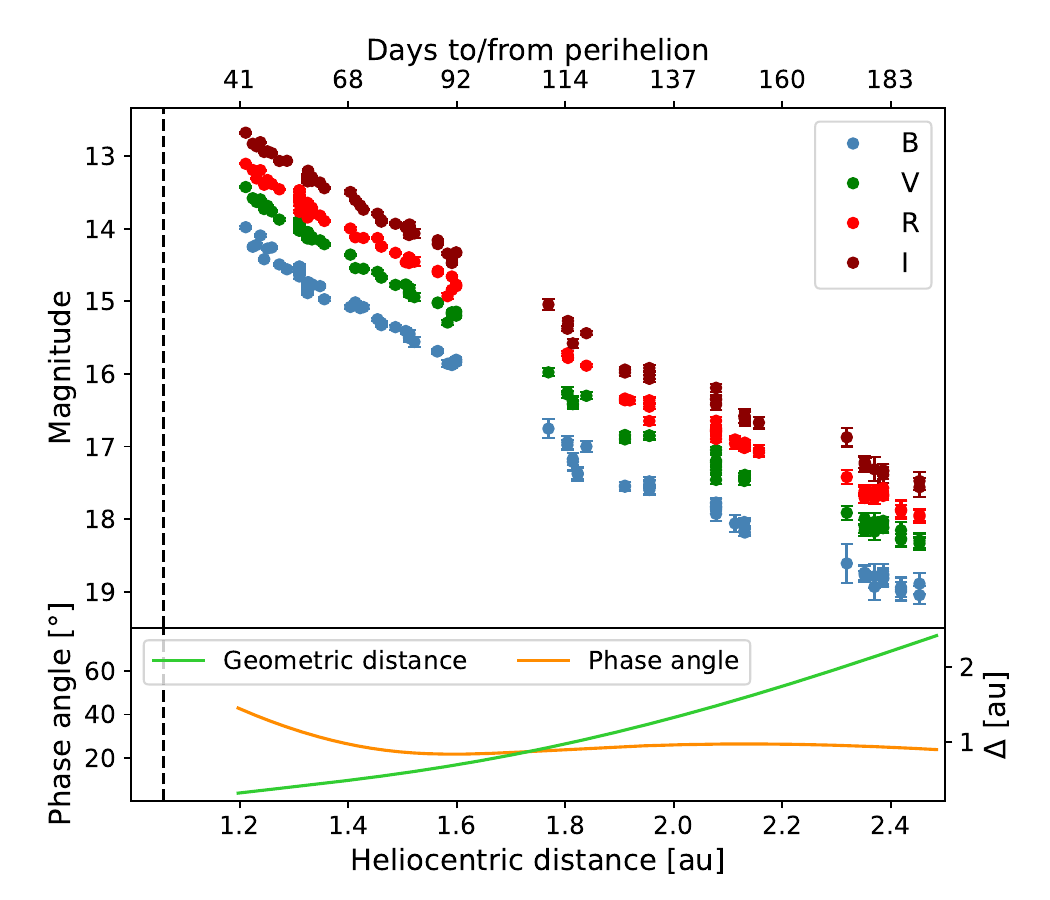} \\
        \includegraphics[width=\linewidth]{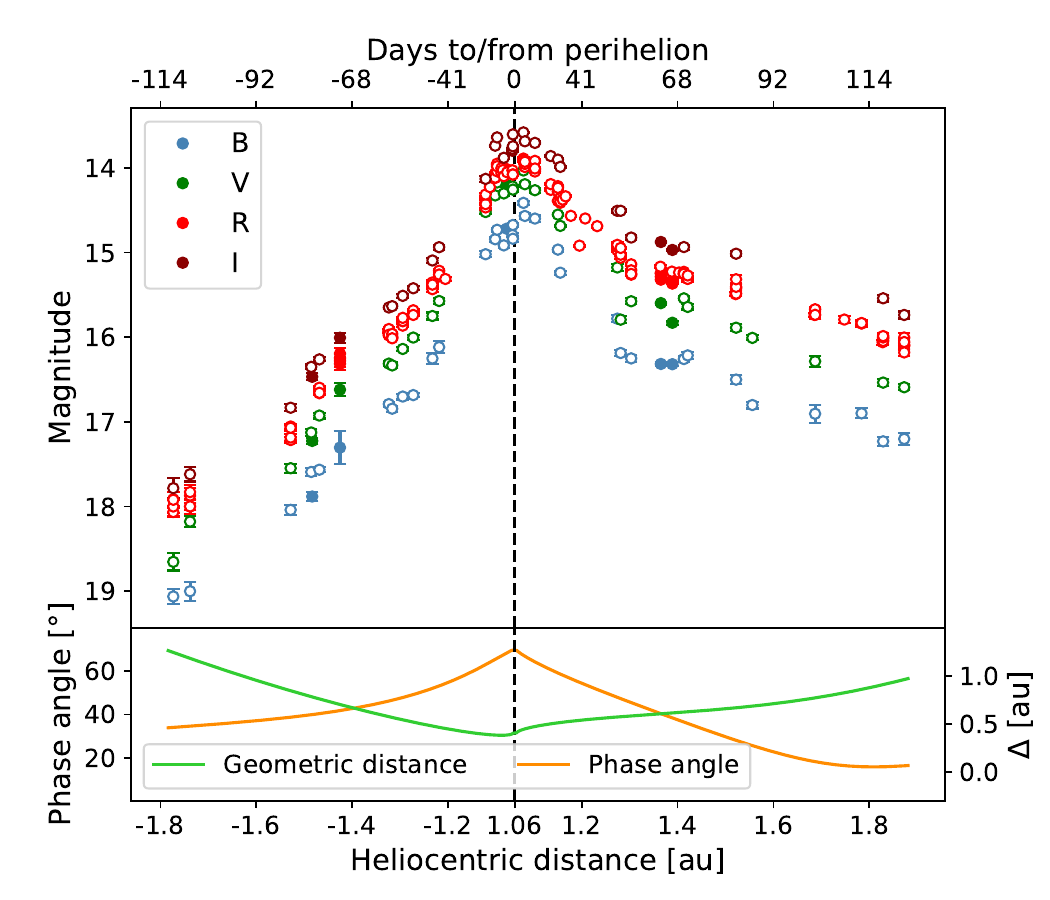}
    \caption{Light curves of comet 103P measured within a 5-arcsec aperture as a function of heliocentric distance and time during its 2010 (top) and 2023 (bottom) perihelion passages. The dotted black line indicates the perihelion at 1.06 au on October 28, 2010, and October 12, 2023.}
    \label{fig:mag_103P}
\end{figure}

The light curves obtained in the different colors for the first and second passages of 67P are shown in Fig. \ref{fig:mag_67P}. Negative heliocentric distances and days stand for the pre-perihelion phase (comet approaching the Sun). Apart from the decrease and increase generally observed around perihelion, we did not detect any outburst in the activity of the comet, contrary to what had been measured in situ by the Rosetta spacecraft during the 2015 perihelion passage (see for example \cite{feldman2016nature} or \cite{2018MNRAS.481.1235R}). Thus, despite the great accuracy reached with the TRAPPIST telescopes, the small, local outbursts occurring inside the coma seem difficult to observe when measuring the global magnitude of comets with ground-based instruments. Fig. \ref{fig:mag_103P} illustrates the lightcurves we obtained for 103P. We restate that only post-perihelion data were available for the 2010 passage. However, the magnitude and therefore the activity of this comet follow the same trend as 67P, as indicated by the 2023 curves. For both comets, we observed a small shift of a few days between the perihelion and the maximum brightness. This result is often observed for comets and could be due either to some seasonal effects linked to the orientation of their spin axis or the repartition of the active sources on their surface.

\begin{figure}[htbp]
    \centering
        \includegraphics[width=\linewidth]{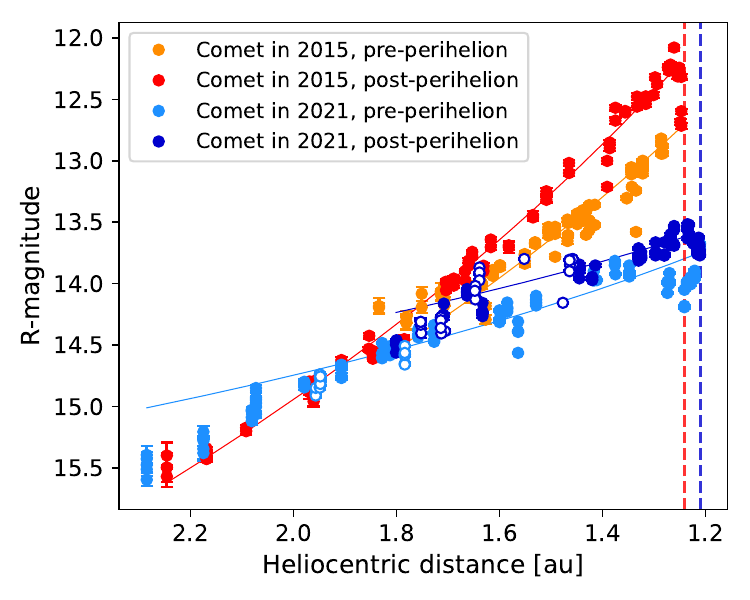} \\
        \includegraphics[width=\linewidth]{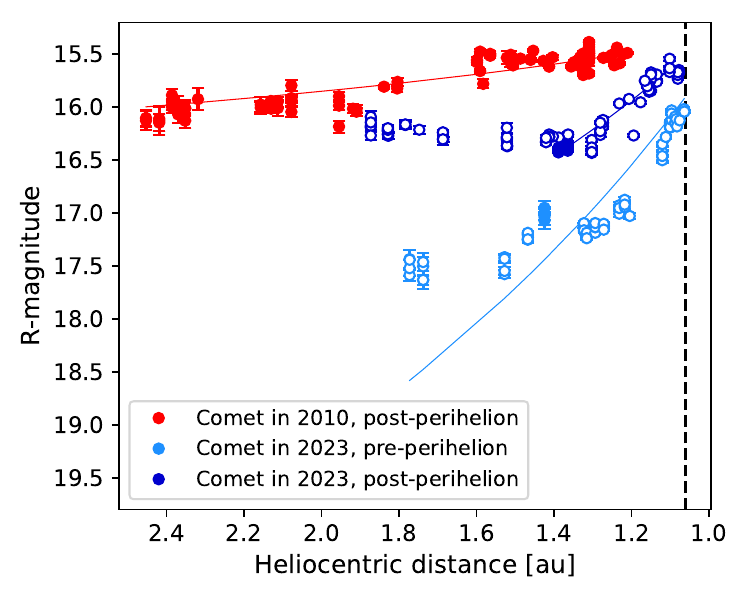}
    \caption{Evolution of the heliocentric magnitude of comets 67P (up) and 103P (bottom) computed with the broadband R filter as a function of heliocentric distance during two perihelion passages. The dotted lines represent the heliocentric distance at perihelion. The colored curves show the fit applied to the datasets. Filled and open points stand for TS and TN data, respectively.}
    \label{fig:heliocentric_mag}
\end{figure}

\begin{table}[h]
\small
\centering
\caption{Absolute magnitude $M$ and power-law exponent $n$ of the magnitude equation fit for the heliocentric magnitude obtained with the R filter.}  
    \begin{tabular}{l >{\centering\arraybackslash}m{1.2cm} >{\centering\arraybackslash}m{1.2cm} >{\centering\arraybackslash}m{1.2cm} >{\centering\arraybackslash}m{1.2cm}}
    \hline
        & \multicolumn{2}{c}{$M$} & \multicolumn{2}{c}{$n$} \\ 
        & \small{Before peak} & \small{After peak} & \small{Before peak} & \small{After peak} \\
    \hline
    \hline
        67P (2015) & 11.64\scriptsize$\pm$0.08 & 10.90\scriptsize$\pm$0.03 & 4.54\scriptsize$\pm$0.20 & 5.37\scriptsize$\pm$0.06 \\
        67P (2021) & 13.37\scriptsize$\pm$0.03 & 13.26\scriptsize$\pm$0.05 & 1.83\scriptsize$\pm$0.06 & 1.52\scriptsize$\pm$0.11 \\
    \hline
        103P (2010) & $-$ & 15.35\scriptsize$\pm$0.02 & $-$ & 0.66\scriptsize$\pm$0.06 \\
        103P (2023) & 15.59\scriptsize$\pm$0.05 & 15.31\scriptsize$\pm$0.04 & 4.80\scriptsize$\pm$0.25 & 3.16\scriptsize$\pm$0.16 \\
    \hline
    \end{tabular}  
\label{tab:fit_mag}
\end{table}

To compare the behavior of comets from one passage to another, it is important to take into account the geometrical configuration of the target with respect to the Sun and Earth, since it can significantly affect the measured light fluxes. The evolution of the geometric distance and the phase angle of the comets during our observations is shown on each plot in Fig. \ref{fig:mag_67P} and \ref{fig:mag_103P}. To illustrate this dependence on geometrical parameters, we computed the heliocentric magnitude of the comets, i.e., their magnitude without the dependence on the distance to Earth. The standard magnitude formula being $m = M + 5 \log\Delta + 2.5 n \log{r_h}$, where $m$ and $M$ are respectively the apparent and absolute magnitudes, $\Delta$ the geocentric distance, $r_h$ the heliocentric distance, and $n$ the power-law exponent, we subtracted the contribution $5 \log\Delta$ from the measured apparent magnitude, as if comets were localized at $\Delta = 1$ au. In this way, the comets' observed brightness only depends on the heliocentric distance, which is the main driver of cometary activity. The $\Delta$ values used for the computation of the heliocentric magnitude were obtained through NASA's JPL Horizons System\footnote{\url{https://ssd.jpl.nasa.gov/horizons/}}.

The results obtained with the R filter are shown in Fig. \ref{fig:heliocentric_mag} for both comets. For 67P, near the perihelion ($r_h < 1.8$ au), we measured a higher (fainter) magnitude of the coma in 2021 than in 2015, indicating a lower overall activity level. This difference gets larger as the comet gets closer to the Sun. We then fit the data before and after the brightness peak with the magnitude formula (without the Earth-distance dependence) to retrieve the $M$ and $n$ parameters. The results of the fit for both passages are overplotted in Fig. \ref{fig:heliocentric_mag} and the $M$ and $n$ values obtained are given in Table \ref{tab:fit_mag}. We observed a higher absolute magnitude in 2021 than in 2015, meaning that 67P was intrinsically brighter and therefore more active during the 2015 passage. The power-law exponent $n$ is also substantially higher in 2015 than in 2021, indicating that, in 2015, the brightness increased much more steeply as the comet approached the Sun. In contrast, the lower $n$ values obtained in 2021 reflect a shallower and more gradual brightening, suggesting that the activity was more spread out across a wider range of heliocentric distances rather than being sharply concentrated near perihelion. This difference between the two apparitions could reflect some changes in the active surface area, the dust-to-gas ratio, or the thermal state of the nucleus, possibly influenced by the significant erosion and resurfacing experienced during the 2015 passage. For 103P, the comparison between the 2010 and 2023 passages is more limited due to the absence of pre-perihelion data for 2010. Nevertheless, the post-perihelion heliocentric magnitudes reveal a striking difference: the 2010 passage was brighter than 2023, particularly for $r_h < 1.9$ au, with the 2010 data points sitting at lower magnitude values. This indicates that 103P was considerably more active in 2010 than in 2023, consistent with a general secular fading of the comet, a phenomenon observed in several short-period comets as they progressively deplete their volatile reservoir with each perihelion passage. The results of the fit are shown in Fig. \ref{fig:heliocentric_mag}, but note that it was restricted to heliocentric distances below 1.4 au, due to a visible change in the slope of the light curve at that distance. This change suggests a transition in the dominant activity regime, possibly marking the growing influence of water ice sublimation as the primary driver as the comet approaches the Sun, replacing more volatile species active at larger distances. The fit $M$ and $n$ values obtained are also given in Table \ref{tab:fit_mag}. The $n$ value for 2023 is notably higher than for 2010, indicating a much steeper brightness decline with increasing heliocentric distance in 2023. The very low $n$ in 2010 suggests an unusually flat fading profile post-perihelion, which may be linked to the exceptionally high CO$_2$-driven activity that the EPOXI mission documented during that apparition \citep{2011ApJ...734L...1M}, sustaining the coma brightness even as the comet receded from the Sun.

The average values (over the two passages) of the 67P slopes, before and after perihelion, are very symmetric, with values of 3.19 $\pm$ 0.10 and 3.45 $\pm$ 0.06, showing a rather homogeneous nucleus in terms of active regions. These values are similar to the pre- and post-perihelion average slopes of 4.5 $\pm$ 1.0 and 2.9 $\pm$ 0.7 found from TRAPPIST R-band measurements by \cite{2023LPICo2851.2093G} for five Jupiter-family comets in the same heliocentric range ($r_h <$ 3 au). With average values of 4.80 $\pm$ 0.25 and 1.91 $\pm$ 0.09 before and after perihelion, 103P slopes are more asymmetrical probably due to some seasonal effects or an inhomogeneous nucleus. Nevertheless, they are again in good agreement with Gardener's values. The slopes of our two Jupiter-family comets are also clearly steeper than the pre- and post-perihelion values found for three long-period comets, respectively 2.1 $\pm$ 1.7 and 1.9 $\pm$ 0.8, and especially for dynamically new comets, which have slope values of only 0.2 $\pm$ 1.6 and 1.5 $\pm$ 1.2 for seven of them. Regardless of the fact that there is some dispersion and that such an analysis should be carried out on more comets, the trends seem clear. This result shows the interest to studying these slopes over high-quality comet light curves, as they seem to reveal different activity behaviors for the various dynamical populations, pointing probably to a different surface composition or geology.

\begin{figure}[htbp]
    \centering
        \includegraphics[width=\linewidth]{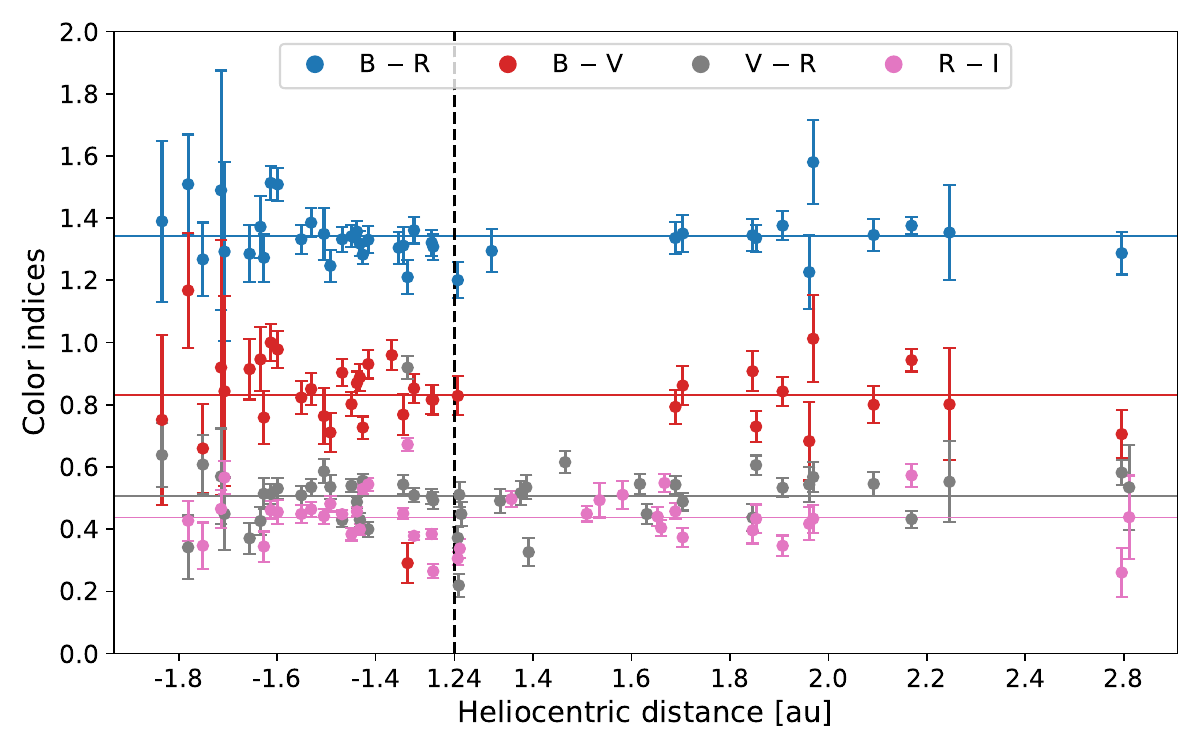} \\
        \includegraphics[width=\linewidth]{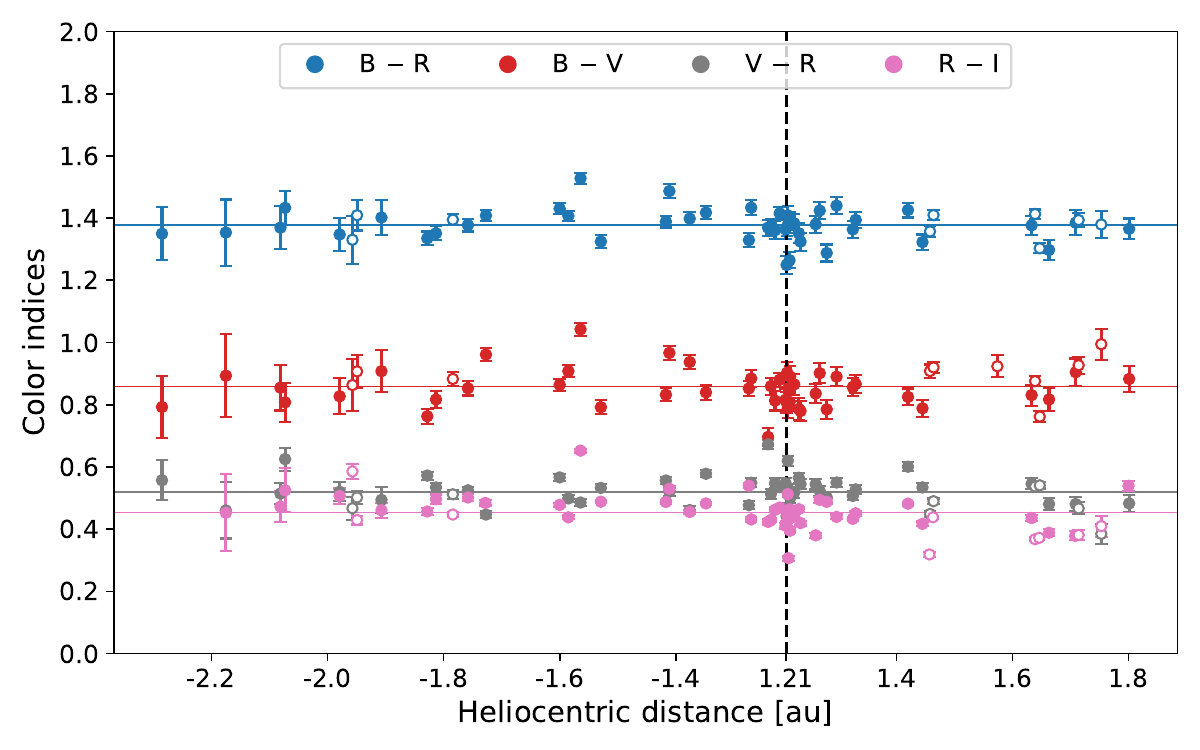}
    \caption{Color indices of comet 67P as a function of heliocentric distance during its 2015 (up) and 2021 (bottom) perihelion passages. The dashed black line represents the heliocentric distance at perihelion. Filled and open points stand for TS and TN data, respectively. The mean value of each color index is represented as a horizontal line.}
    \label{fig:color_67P}
\end{figure}

\begin{figure}[htbp]
    \centering
        \includegraphics[width=\linewidth]{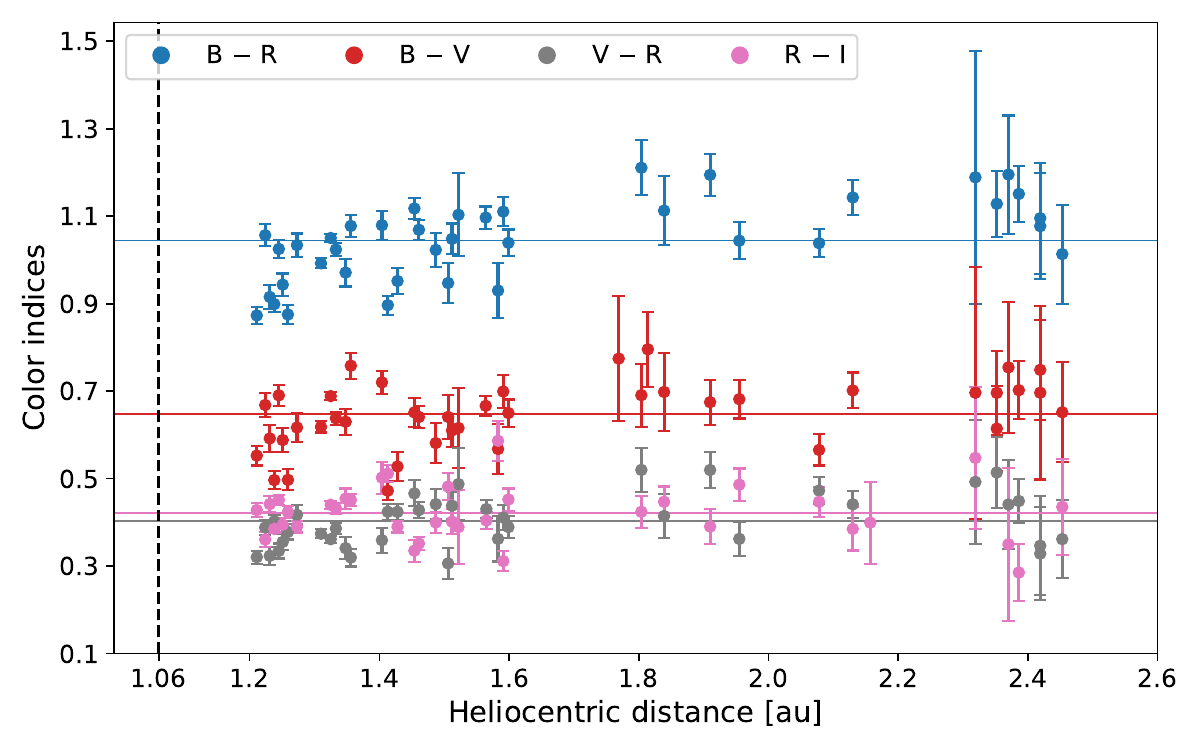} \\
        \includegraphics[width=\linewidth]{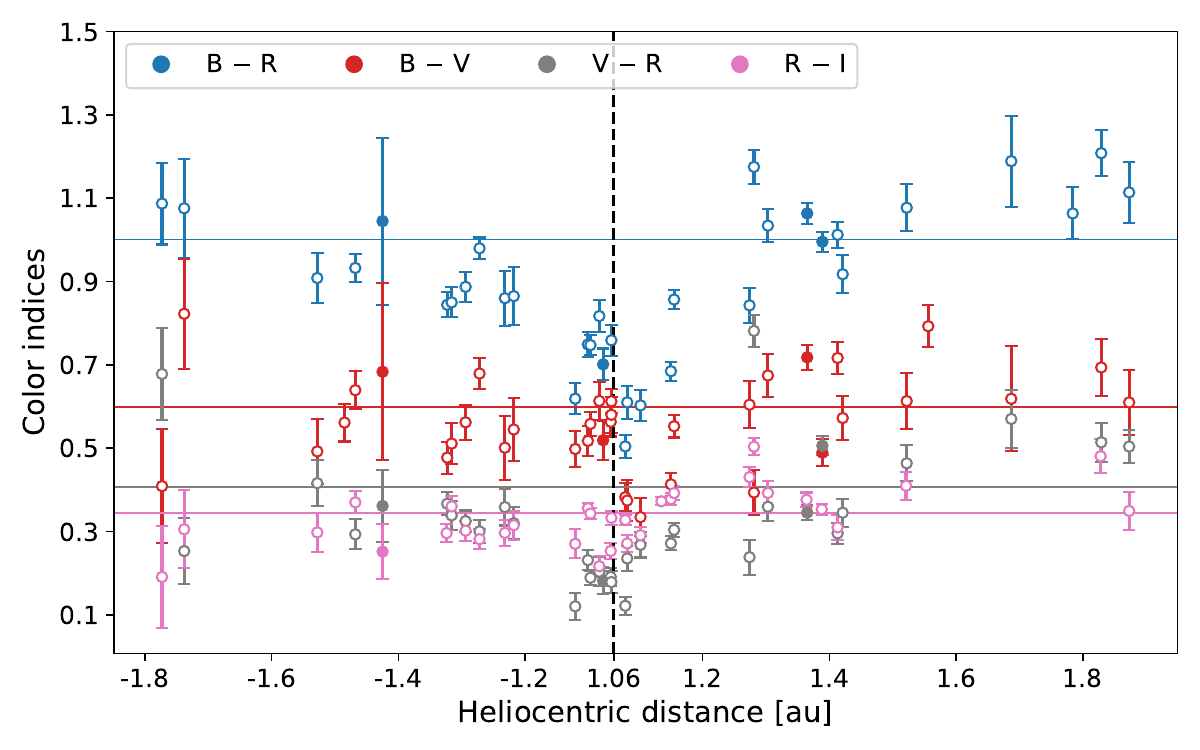}
    \caption{Color indices of comet 103P as a function of heliocentric distance during its 2010 (up) and 2023 (bottom) perihelion passages.}
    \label{fig:color_103P}
\end{figure}

The study of the colors of comets' coma, which can provide some information about dust grains, was done through the computation of B$-$R, B$-$V, V$-$R, and R$-$I color indices. The evolution of these indices during the first and second passages of 67P and 103P is shown in Fig. \ref{fig:color_67P} and \ref{fig:color_103P}, respectively. Our analysis does not reveal any significant variation of the colors (and therefore the dust grain properties) throughout the heliocentric distance range, except near the perihelion during the second passage of 103P, where the B$-$R indices are decreasing up to 0.6 and then increasing again, which was most probably due to contamination of the B filter by the CN gas emission. The mean value of each color index is given in Table \ref{tab:colors} for both comets. This table also gives the mean values of the color indices for active Jupiter-family comets computed by \cite{2012Icar..218..571S}, those for active long-period comets computed by \cite{2015AJ....150..201J}, and the color indices of the Sun given in \cite{holmberg2006colours}. For 67P, we did not observe significant differences in the colors between 2015 and 2021. Our values are very close to the mean indices computed for Jupiter-family comets and are higher than the indices of the Sun, which indicates that 67P's colors are redder than the solar colors. For 103P, we computed the mean indices for the 2023 passage without taking the values obtained at $r_h < 1.2$ au into account, to get a more accurate comparison with the 2010 indices, since we did not have data for smaller heliocentric distances during the first passage. Thus, for heliocentric distances larger than 1.2 au, the color indices of the comet remained almost constant between 2010 and 2023. Moreover, for both passages, our values are highly consistent with the mean values for Jupiter-family comets. Finally, we also observed that, as for 67P, the comet colors are redder than the solar colors. 

\begin{table}
\centering
\caption{Mean color indices computed for comets 67P and 103P during both perihelion passages and comparison with colors of active Jupiter-family comets (JFCs) and long-period comets (LPCs) and solar colors.}
\begin{threeparttable}
\small
    \begin{tabular}{l c c c c}
        \hline
            Object & B $-$ R & B $-$ V & V $-$ R & R $-$ I \\     
        \hline
        \hline
            67P (2015) & 1.34\scriptsize$\pm$0.02 & 0.83\scriptsize$\pm$0.03 & 0.51\scriptsize$\pm$0.02 & 0.44\scriptsize$\pm$0.02 \\
            67P (2021) & 1.38\scriptsize$\pm$0.01 & 0.86\scriptsize$\pm$0.01 & 0.52\scriptsize$\pm$0.01 & 0.45\scriptsize$\pm$0.01 \\
            Evolution & +0.04\scriptsize$\pm$0.02 & +0.03\scriptsize$\pm$0.03 & +0.01\scriptsize$\pm$0.02 & +0.01\scriptsize$\pm$0.02 \\
        \hline
            103P (2010) & 1.04\scriptsize$\pm$0.01 & 0.65\scriptsize$\pm$0.01 & 0.40\scriptsize$\pm$0.01 & 0.42\scriptsize$\pm$0.01 \\
            103P (2023) & 1.00\scriptsize$\pm$0.02 & 0.60\scriptsize$\pm$0.02 & 0.41\scriptsize$\pm$0.03 & 0.34\scriptsize$\pm$0.02 \\
            Evolution & -0.04\scriptsize$\pm$0.02 & -0.05\scriptsize$\pm$0.02 & +0.01\scriptsize$\pm$0.03 & -0.08\scriptsize$\pm$0.02 \\
        \hline
        \hline
            Active JFCs\tnote{a} & 1.22\scriptsize$\pm$0.02 & 0.75\scriptsize$\pm$0.02 & 0.47\scriptsize$\pm$0.02 & 0.44\scriptsize$\pm$0.02 \\            
            Active LPCs\tnote{b} & 1.24\scriptsize$\pm$0.02 & 0.78\scriptsize$\pm$0.02 & 0.47\scriptsize$\pm$0.02 & 0.42\scriptsize$\pm$0.03 \\            
            Sun\tnote{c} & 0.99\scriptsize$\pm$0.02 & 0.64\scriptsize$\pm$0.02 & 0.35\scriptsize$\pm$0.01 & 0.33\scriptsize$\pm$0.01 \\
        \hline
    \end{tabular}
    \begin{tablenotes}
        \item[a] Values from \cite{2012Icar..218..571S}.
        \item[b] Values from \cite{2015AJ....150..201J}.
        \item[c] Values from \cite{holmberg2006colours}.
    \end{tablenotes}
\end{threeparttable}    
\label{tab:colors}
\end{table}

\subsection{Computation of gas production rates}

Based on the data obtained from TRAPPIST, we computed the production rates of the molecules OH, NH, CN, C$_2$, and C$_3$ in the coma by converting the flux measured within their corresponding wavelength range (defined by the bandpass of the narrowband filters) into column densities and fitting a Haser model \citep{1957BSRSL..43..740H} to the resulting radial column density profiles. For the spectroscopic data obtained from HCT, the flux of the molecules CN, C$_2$, and C$_3$ within their corresponding wavelength range, extracted from equivalent apertures in the spatial axis, was converted to column densities. The resulting column density profiles as a function of nucleocentric distance were then fit with the Haser model using the minimum chi-square estimation to compute the corresponding production rate.

In both cases, the Haser model has been adjusted to our median radial profiles for nucleocentric distances between $10^{3.6}$ and $10^{4.1}$ km. This range was chosen as a compromise between seeing effects and dust contamination that can significantly affect the images at a small nucleocentric distance and the too low signal-to-noise ratio obtained at a large distance. For the fluorescence efficiencies (also known as g-factors) and scale lengths of the parent and daughter species, necessary for the conversion of fluxes into column densities, we used values from \cite{1995Icar..118..223A}.

\begin{figure*}[htbp]
    \centering
    \begin{subfigure}{0.48\textwidth}
        \centering
        \includegraphics[width=\linewidth]{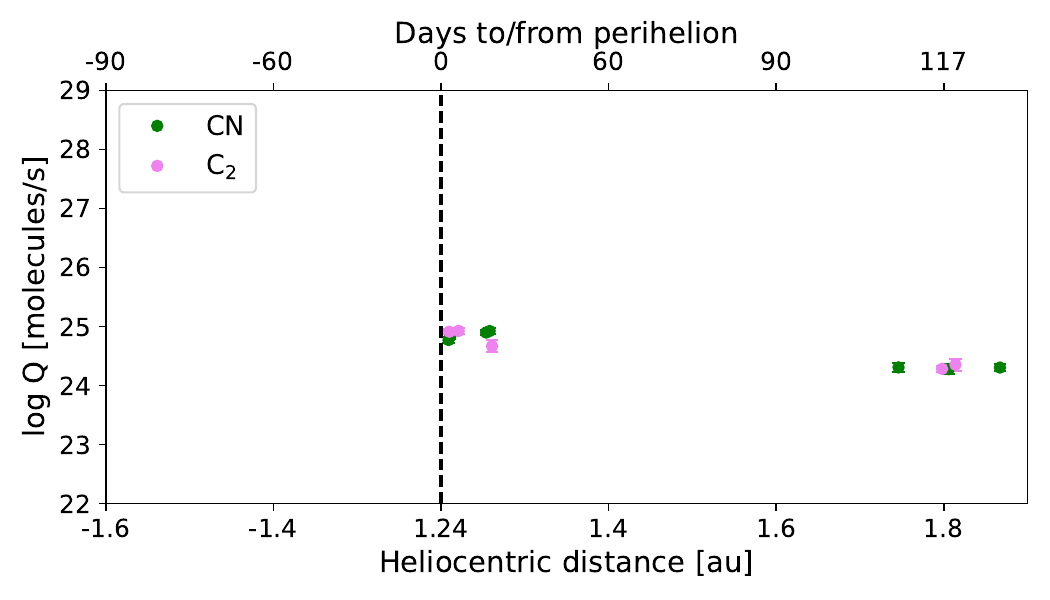}
    \end{subfigure}
    \hfill
    \begin{subfigure}{0.48\textwidth}
        \centering
        \includegraphics[width=\linewidth]{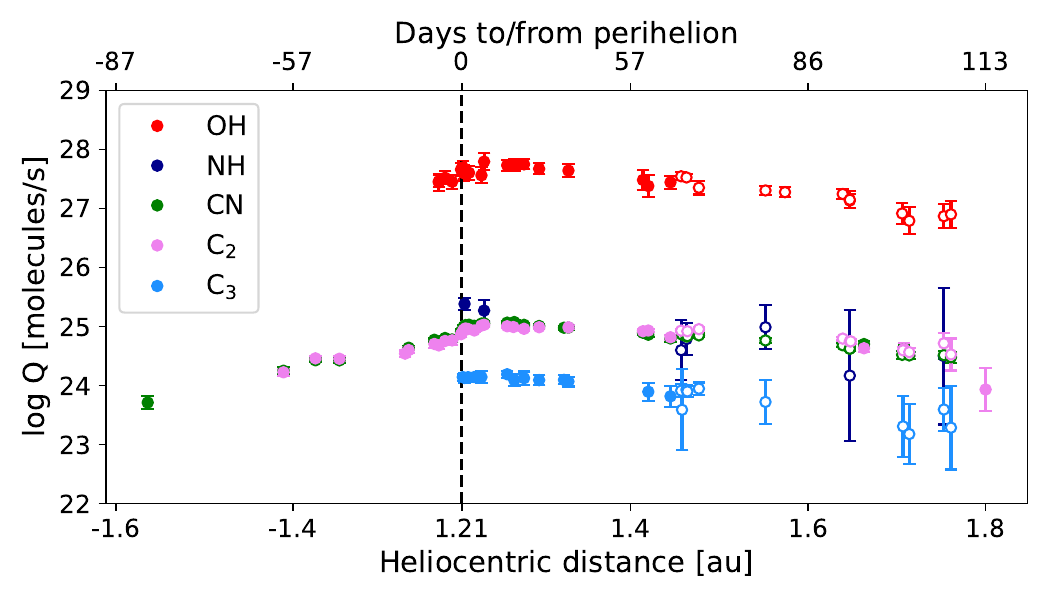}
    \end{subfigure}
    \caption{Evolution of the molecular production rates of comet 67P as a function of the heliocentric distance and time during its 2015 (left) and 2021 (right) apparitions. The dotted line indicates the perihelion. Filled and open points stand for TS and TN data, respectively.}
    \label{fig:Q_all_0067P}
\end{figure*}

\begin{figure*}[htbp]
    \centering
    \begin{subfigure}{0.48\textwidth}
        \centering
        \includegraphics[width=\linewidth]{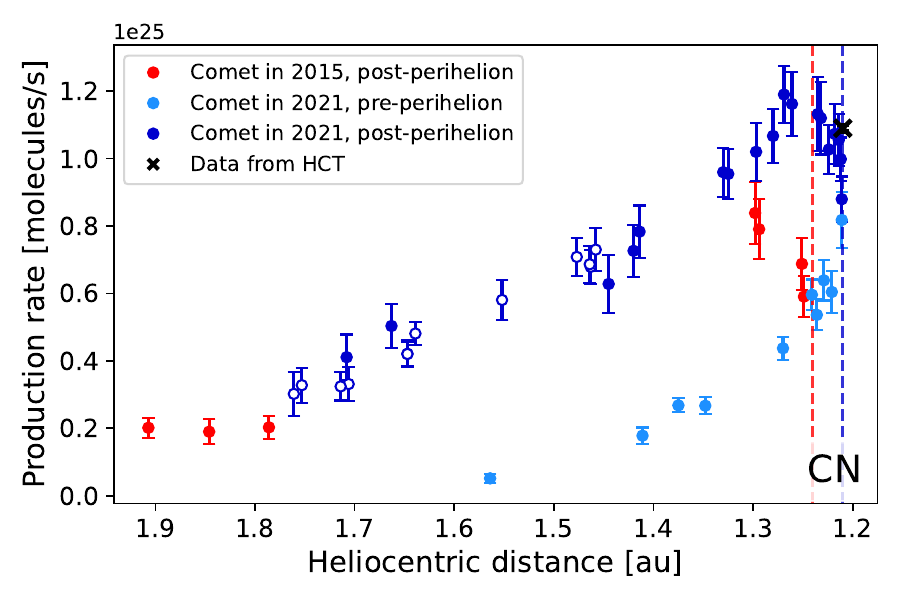}
    \end{subfigure}
    \hfill
    \begin{subfigure}{0.48\textwidth}
        \centering
        \includegraphics[width=\linewidth]{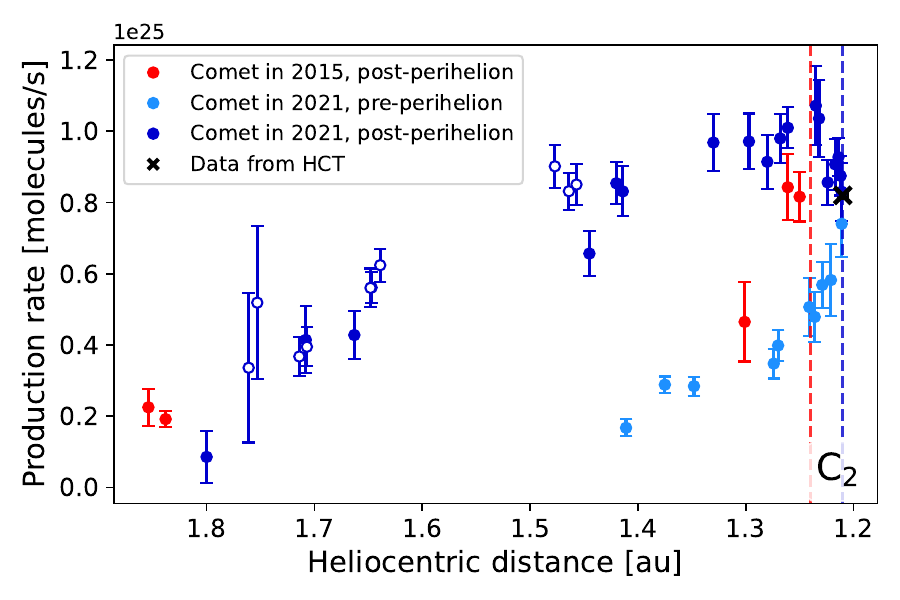}
    \end{subfigure}
    \caption{Evolution of the CN and C$_2$ production rates of comet 67P as a function of the heliocentric distance. The black cross indicates the production rate obtained through spectroscopic analysis with HCT on November 9, 2021.}
    \label{fig:Q_CN_C2_0067P}
\end{figure*}

\begin{figure*}[htbp]
    \centering
    \begin{subfigure}{0.48\textwidth}
        \centering
        \includegraphics[width=\linewidth]{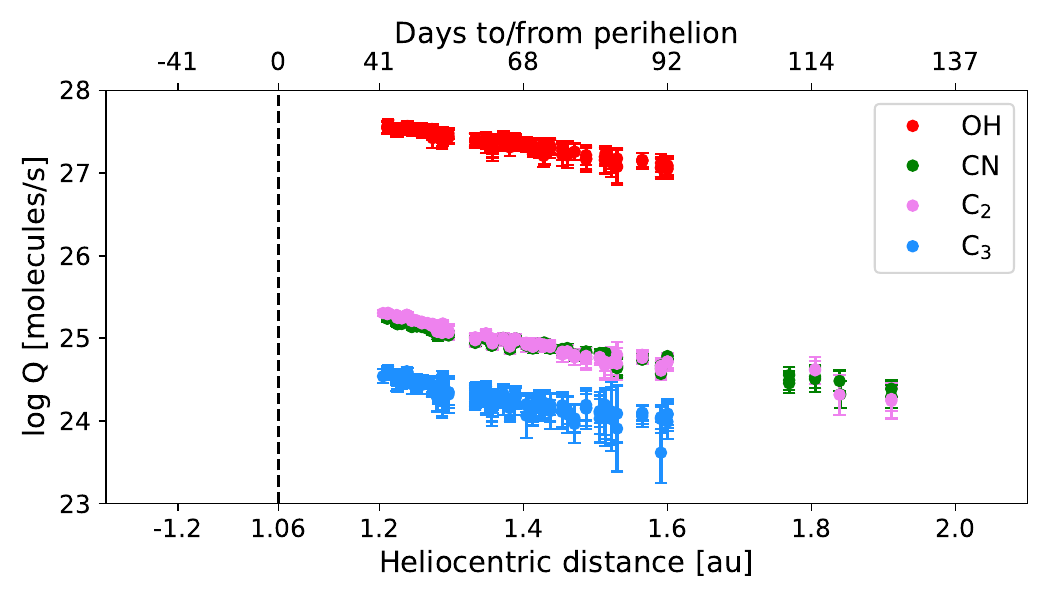}
    \end{subfigure}
    \hfill
    \begin{subfigure}{0.48\textwidth}
        \centering
        \includegraphics[width=\linewidth]{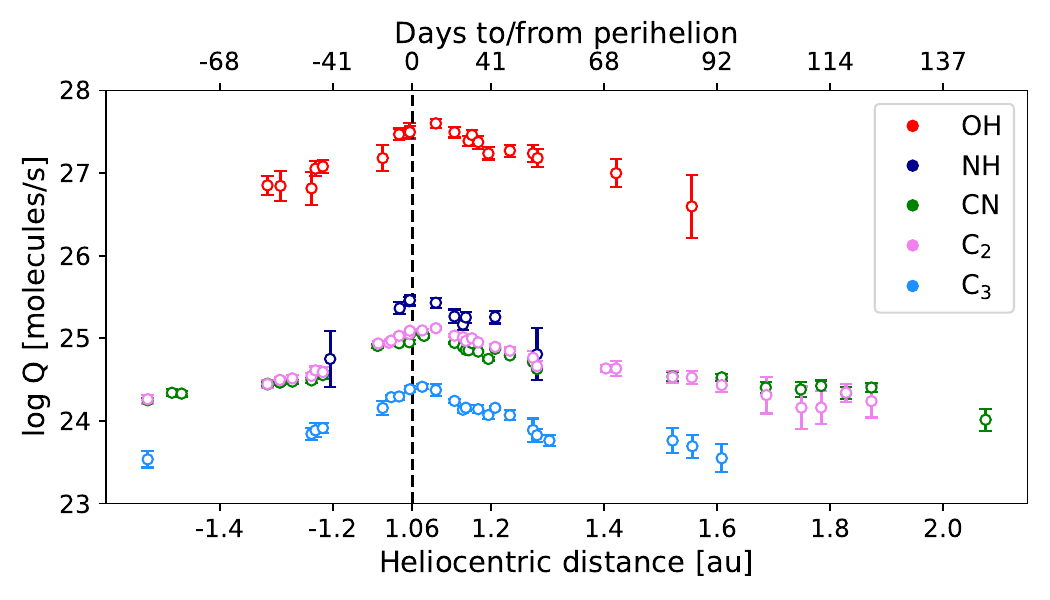}
    \end{subfigure}
    \caption{Evolution of the molecular production rates of comet 103P as a function of the heliocentric distance and time during its 2010 (left) and 2023 (right) apparitions. The dotted line indicates the perihelion. Filled and open points stand for TS and TN data, respectively.}
    \label{fig:Q_all_0103P}
\end{figure*}

\begin{figure*}[htbp]
    \centering
    \begin{subfigure}{0.48\textwidth}
        \centering
        \includegraphics[width=\linewidth]{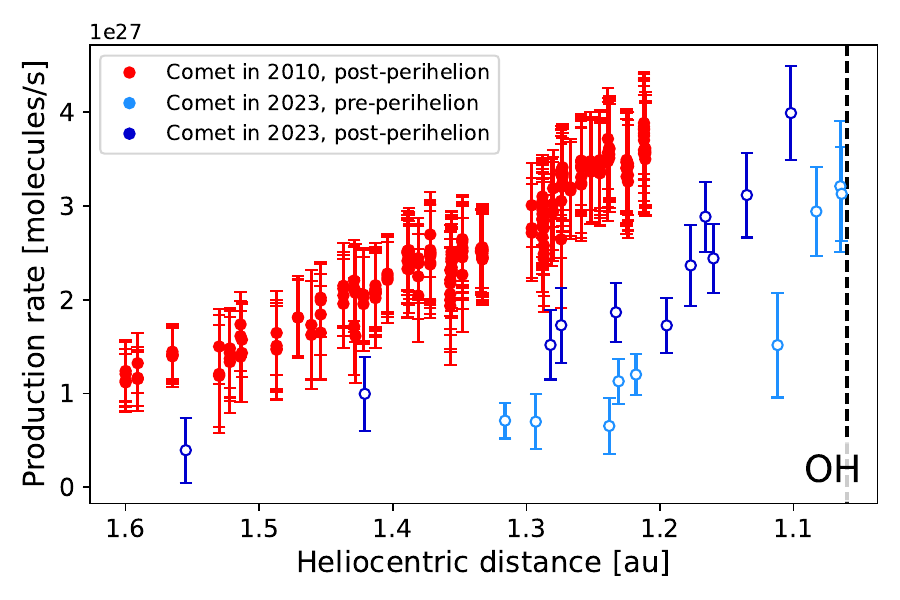}
    \end{subfigure}
    \hfill
    \begin{subfigure}{0.48\textwidth}
        \centering
        \includegraphics[width=\linewidth]{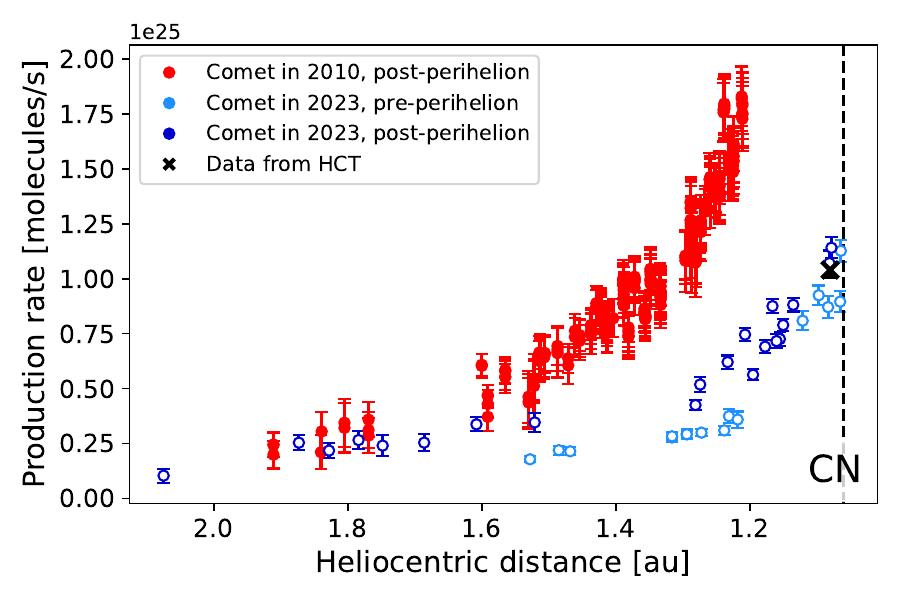}
    \end{subfigure}
    \begin{subfigure}{0.48\textwidth}
        \centering
        \includegraphics[width=\linewidth]{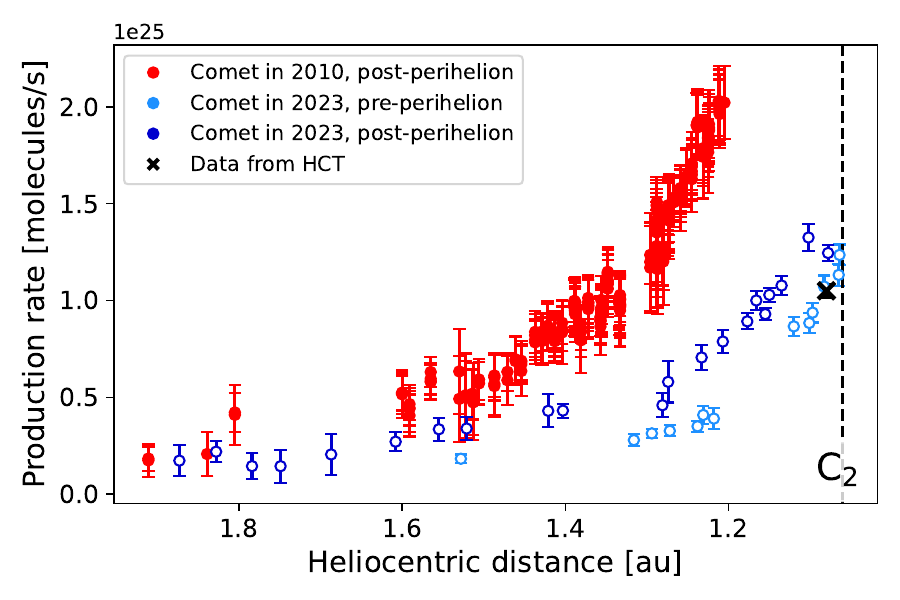}
    \end{subfigure}
    \hfill
    \begin{subfigure}{0.48\textwidth}
        \centering
        \includegraphics[width=\linewidth]{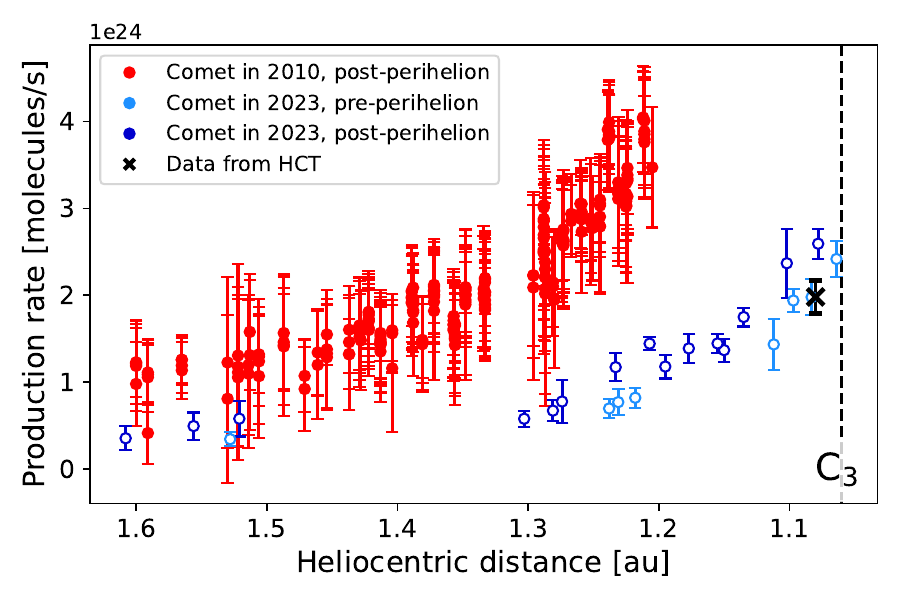}
    \end{subfigure}
    \caption{Evolution of the OH, CN, C$_2$, and C$_3$, production rates of comet 103P as a function of the heliocentric distance. The black cross on the C$_2$, C$_3$, and CN plots indicates the production rate obtained through spectroscopic analysis with HCT on September 28, 2023.}
    \label{fig:Q_OH_CN_C2_C3_103P}
\end{figure*}

The OH, NH, CN, C$_2$, and C$_3$ production rates measured with TRAPPIST for the two observed apparitions of comet 67P are shown in Fig. \ref{fig:Q_all_0067P}. In addition, Fig. \ref{fig:Q_CN_C2_0067P} shows the comparison of the CN and C$_2$ production rates measured in 2015 and in 2021. Note that for the OH, NH, and C$_3$ gaseous species, data were not available for the 2015 passage, so we could not establish any comparison for the production rates of these gases from one passage to the other. Looking at the data from 2021, we were able to track the evolution of gas production during the comet's passage near the sun, and we observed a fast increase in production rates as the comet approaches the Sun, followed by a decrease as it moves away. The maximum gas production rates is reached between 20 and 50 days after perihelion, depending on the species. The CN and C$_2$ production rates obtained on November 9, 2021, through spectroscopic analysis with HCT are illustrated in the plots of Fig. \ref{fig:Q_CN_C2_0067P} and are in very good agreement with our TRAPPIST measurements. The 2015 data (post-perihelion) seem to follow the same trend, although they are slightly lower than those for 2021. \cite{2016MNRAS.462S.138S} have analyzed the post-perihelion CN production rates of comet 67P during its 2015 passage by collecting data with different ground-based telescopes, including TS. Comparing their results with ours allowed us to overcome the problem of insufficient data. In particular, their measurements performed with the LOw-cosT Ultraviolet Spectrograph (LOTUS) instrument on the Liverpool Telescope are in good agreement with those realized with TS. Those observations provided a more complete dataset that enables a more accurate analysis of the evolution of the CN production rate of the comet in 2015 and confirms that this rate is indeed lower than the one measured in 2021. 

The gas production rates of comet 103P measured through photometry with TRAPPIST during the 2010 and 2023 perihelion passages are shown in Fig. \ref{fig:Q_all_0103P}. In addition, Fig. \ref{fig:Q_OH_CN_C2_C3_103P} shows the difference in the OH, CN, C$_2$, and C$_3$ production rates measured during the two apparitions of the comet. Note that the NH filter was not used for the 2010 observations. Here, we still observed an increase in the activity before perihelion and then a decrease starting around two weeks after perihelion. Although we can not compare our results for the pre-perihelion phase, we unambiguously observed a higher production rate in 2010 as compared to 2023, at least for heliocentric distances higher than 1.2 au in the post-perihelion phase. This difference is more pronounced for C$_2$, C$_3$, and CN and seems to soften at large heliocentric distances. As for 67P, we compared our results with those obtained with HCT. In particular, the CN, C$_2$, and C$_3$ production rates measured through spectroscopy on September 28, 2023, are shown in Fig. \ref{fig:Q_OH_CN_C2_C3_103P} and are in good agreement with TRAPPIST data. 

\begin{figure}[htbp]
    \centering
        \includegraphics[width=\linewidth]{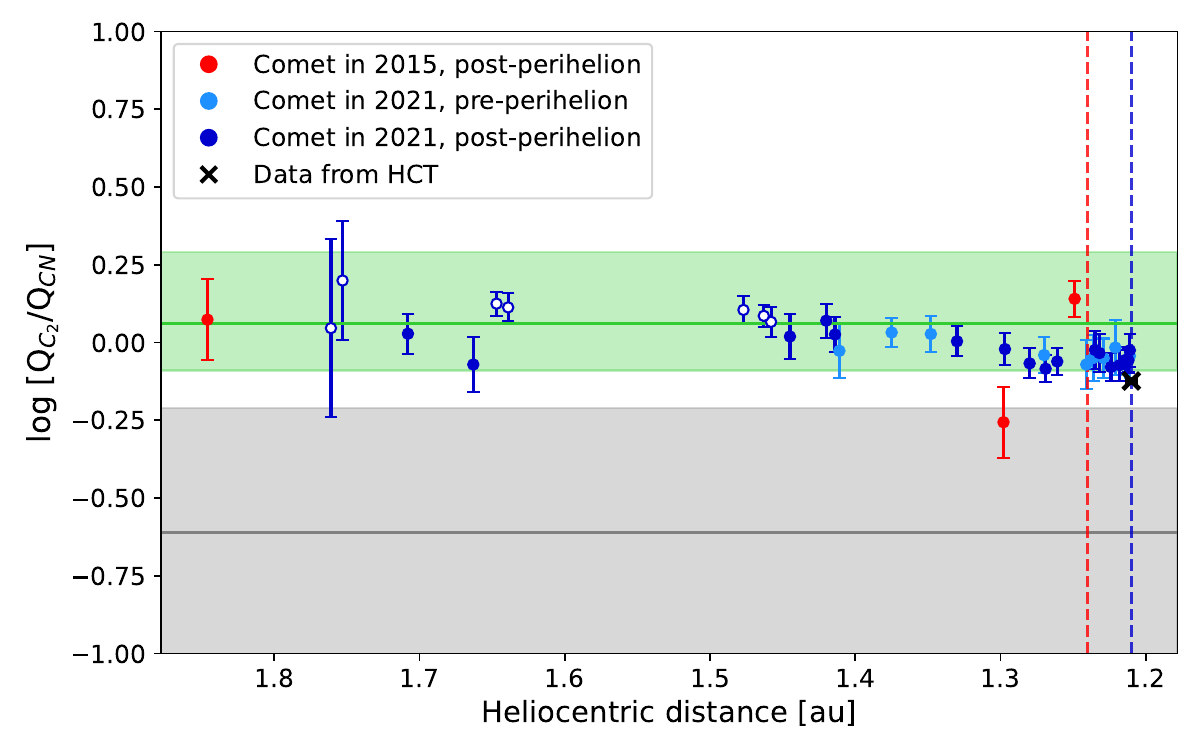} \\
        \includegraphics[width=\linewidth]{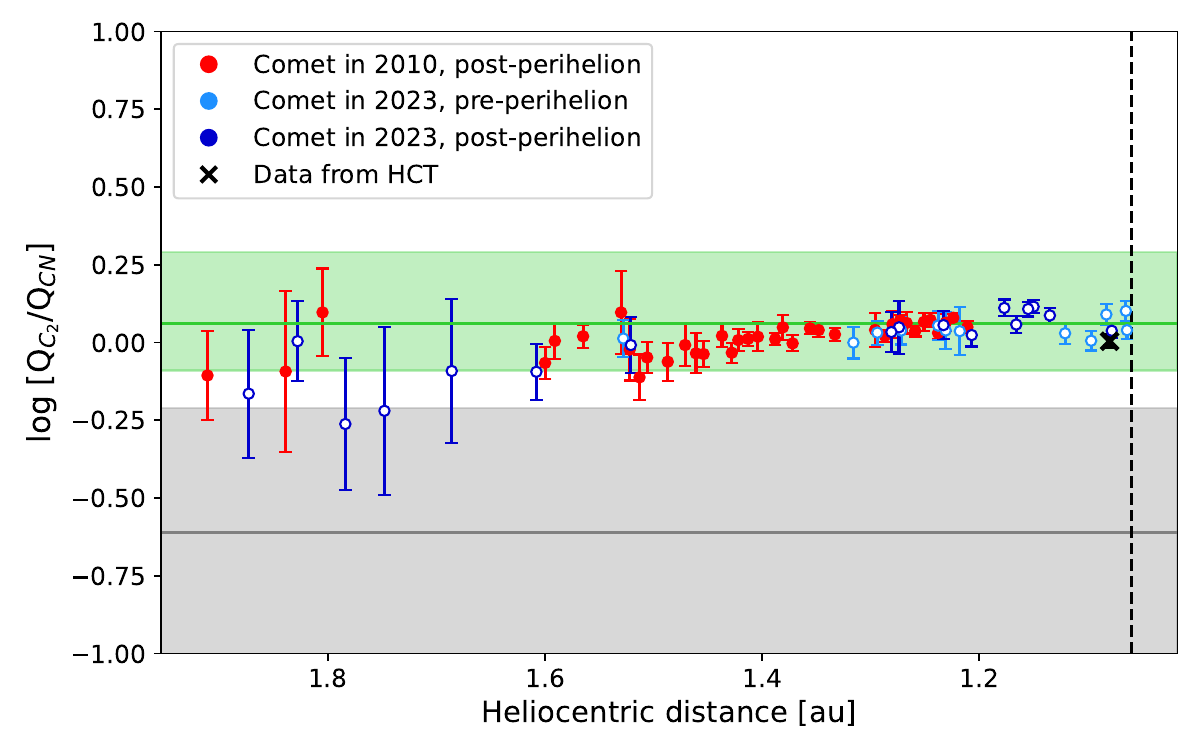}
    \caption{Evolution of the C$_2$-to-CN production rate ratios of comets 67P (top) and 103P (bottom) as a function of the heliocentric distance. The black cross indicates the ratio obtained through spectroscopic analysis with HCT.}
    \label{fig:logQ_C2/CN}
\end{figure}

From the C$_2$ and CN production rates computed for both comets, we also derived ratios in order to determine their chemical class. Indeed, a commonly used taxonomy system consists of dividing comets into two categories based on their chemical composition: the typical ones and those displaying a depletion in species containing carbon chains (C$_2$ and C$_3$). According to the criterion defined by \cite{1995Icar..118..223A}, depleted comets have a C$_2$-to-CN production rate ($Q$) ratio such that $\log[Q_{C_2}/Q_{CN}]<-0.18$, while for the typical comets, the mean value of $\log[Q_{C_2}/Q_{CN}]$ is around $+0.06$. Fig. \ref{fig:logQ_C2/CN} shows the C$_2$-to-CN production rate ratios that we obtained for the two perihelion passages of both comets. For 67P, we observed through the 2021 data a slight variation of the ratios with heliocentric distance. Indeed, they seem to follow a decreasing trend in the pre-perihelion phase starting from around 1.4 au, with a minimum value close to the perihelion, and then they increase in the post-perihelion phase. On the contrary, the production rate ratios of 103P slightly increase during the pre-perihelion phase to reach a maximum value that is very close to the mean value for typical comets, and then decrease after perihelion. For both 67P and 103P, we did not notice any significant variation in the ratio values from one passage to the other, although this result is not very clear for 67P, given the limited data available in 2015. In addition, in this figure, the green area represents the range of ratio values ($-0.09 \longrightarrow 0.29$) obtained by \cite{1995Icar..118..223A} for typical comets, with the green horizontal line showing the mean value ($0.06 \pm 0.10$), and the gray area represents the range of values ($-1.22 \longrightarrow -0.21$) they obtained for depleted comets, with the gray horizontal line showing the mean value ($-0.61 \pm 0.35$). Thus, according to A'Hearn's criterion, both comets can be classified as typical in terms of abundance in carbon-chain species. 

\subsection{Estimation of the dust activity}

The quantification of dust activity of comets 67P and 103P was performed through the computation of a parameter known as $Af\rho$, first introduced by \cite{1984AJ.....89..579A}. Using this parameter, the cometary dust activity can be approximated by the following relation:
\begin{equation*}
    A(\theta)f\rho = \frac{(2\Delta r_h)^2}{\rho}\frac{F_{comet}}{F_{sun}},
\end{equation*} 
where $\Delta$ and $r_h$ are, respectively, the geocentric distance (in cm) and heliocentric distance (in au) of the comet, $F_{comet}$ is the solar flux reflected in the field of view by the dust grains present within the nucleocentric sphere of radius $\rho$, and $F_{sun}$ is the solar flux measured at 1 au. The proxy $A(\theta)f\rho$ is given in cm.

The obtained $A(\theta)f\rho$ values were corrected for the phase angle effect (i.e., their dependence on the phase angle at the time of the observations). This correction is necessary for comparing the dust activity of different comets, or in the case of this work, during different perihelion passages. In practice, we computed an $A(\theta)f\rho$ radial profile for each image obtained with the narrowband dust filters BC, GC, and RC, as well as with the broadband filter R. We then interpolated the profiles at a nucleocentric distance of 10,000 km to obtain a dust activity proxy, and we normalized this proxy at a phase angle $\theta=0$° using the composite dust phase function defined by D. Schleicher\footnote{\url{https://asteroid.lowell.edu/comet/dustphase.html}}. The phase angle of the comets during the observations is illustrated in Fig. \ref{fig:mag_67P} and \ref{fig:mag_103P}.

\begin{figure}[htbp]
    \centering
        \includegraphics[width=\linewidth]{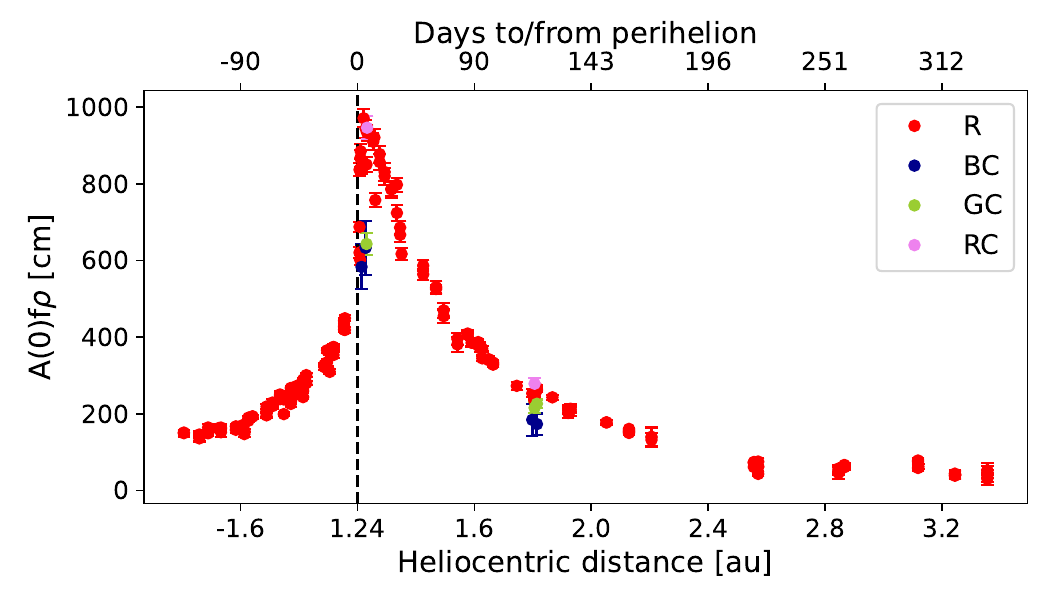} \\
        \includegraphics[width=\linewidth]{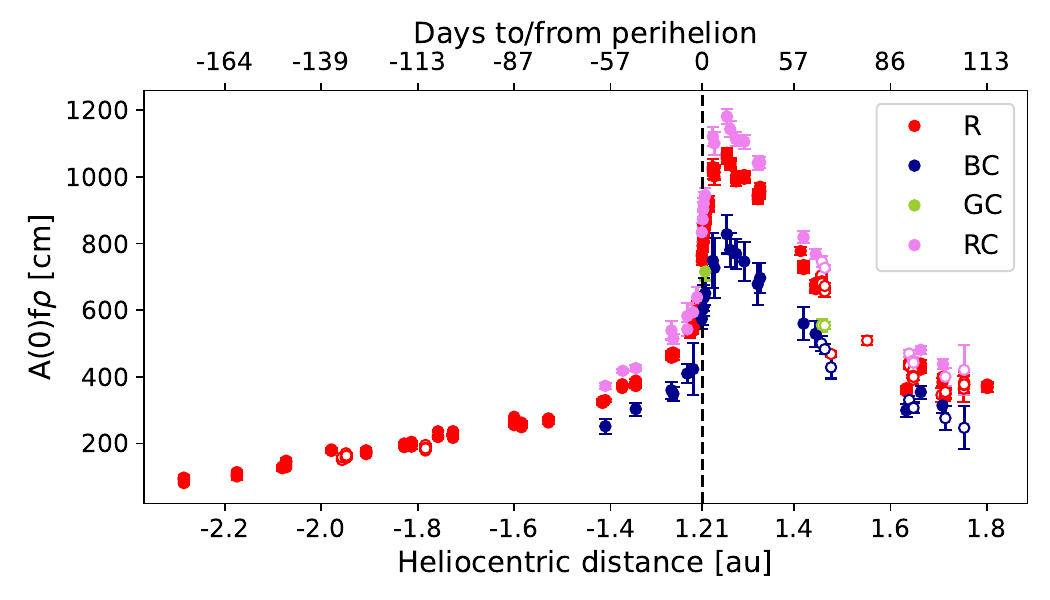}
    \caption{Evolution of the $A(0)f\rho$ parameter of comet 67P as a function of heliocentric distance and time during its 2015 (up) and 2021 (bottom) returns. The dotted black line indicates the perihelion. Filled and open points stand for TS and TN data, respectively.}
    \label{fig:afrho_67P}
\end{figure}

\begin{figure}[htbp]
    \centering
        \includegraphics[width=\linewidth]{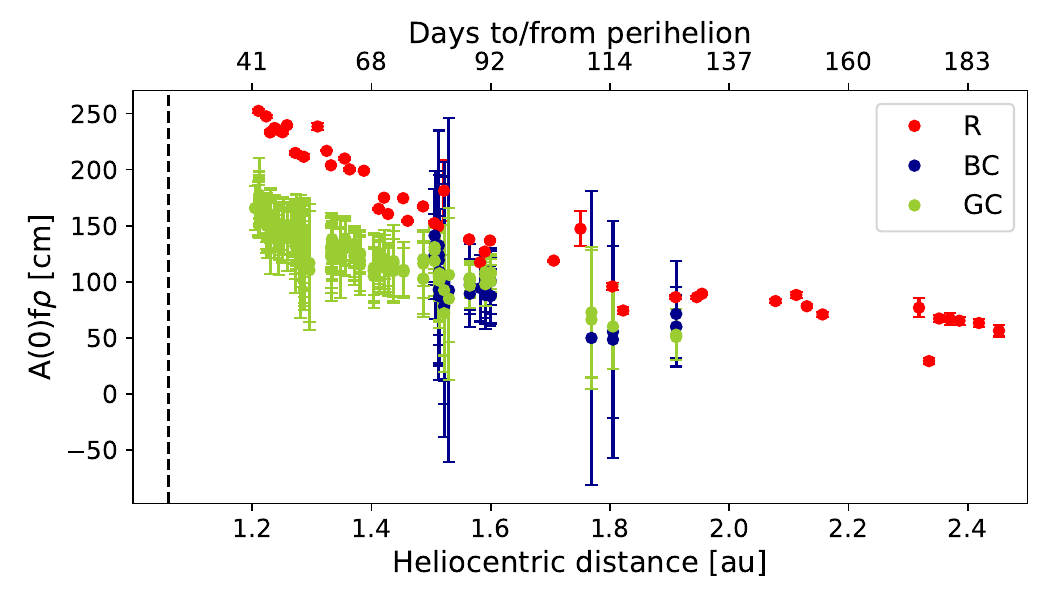} \\
        \includegraphics[width=\linewidth]{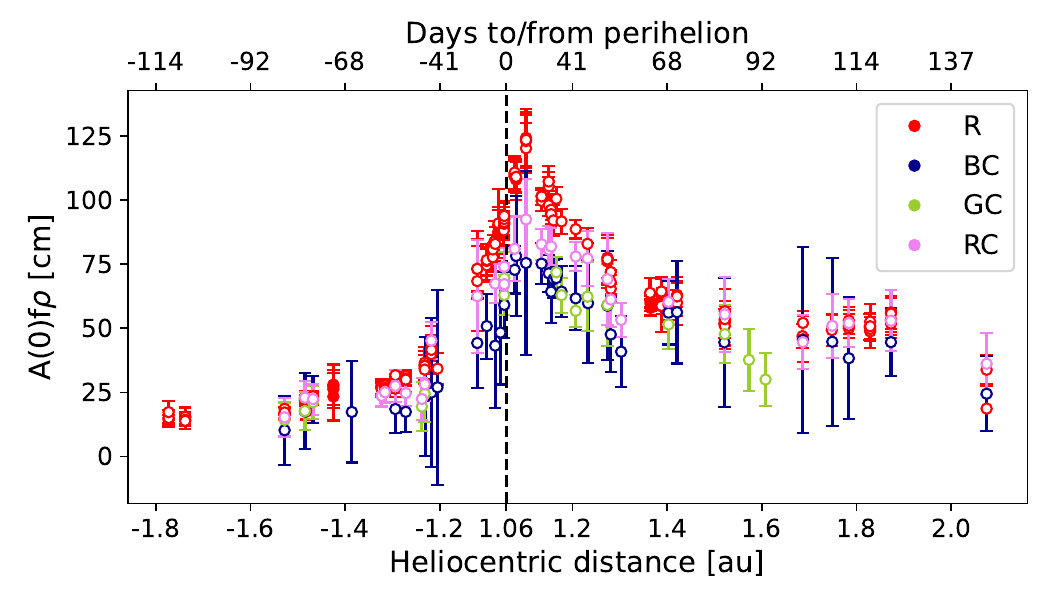}
    \caption{Evolution of the $A(0)f\rho$ parameter of comet 103P as a function of heliocentric distance and time during its 2010 (up) and 2023 (bottom) returns.}
    \label{fig:afrho_103P}
\end{figure}

The $A(\theta=0)f\rho$ values computed for 67P with the narrowband BC, GC, and RC filters and the broadband R filter are illustrated in Fig. \ref{fig:afrho_67P} for the two passages of the comet. We observed a slow increase in the dust activity in the pre-perihelion phase up to approximately 1.6 au, and then the $A(0)f\rho$ values increased much faster until the comet reached peak activity. In the post-perihelion phase, the dust activity decreases as the comet goes away from the Sun, rapidly at first and then more slowly beyond 1.6 au. Maximum activity is detected approximately two weeks after perihelion in 2015, and almost 30 days after perihelion in 2021. The results obtained for 103P are shown in Fig. \ref{fig:afrho_103P}. Note that the RC filter was not used for the observations during the first passage of the comet. The 2010 data give information only about the post-perihelion phase, and we observed a decreasing dust activity from the beginning to the end of our observations. However, the 2023 data show a trend similar to that observed for 67P, i.e., a slowly increasing activity in the pre-perihelion phase up to approximately 1.4 au, followed by a steeper increase until the maximum activity observed around three weeks after perihelion. Then, the $Af\rho$ parameter decreases quickly up to 1.4 au and then much more slowly. 

\begin{figure}[htbp]
    \centering
        \includegraphics[width=\linewidth]{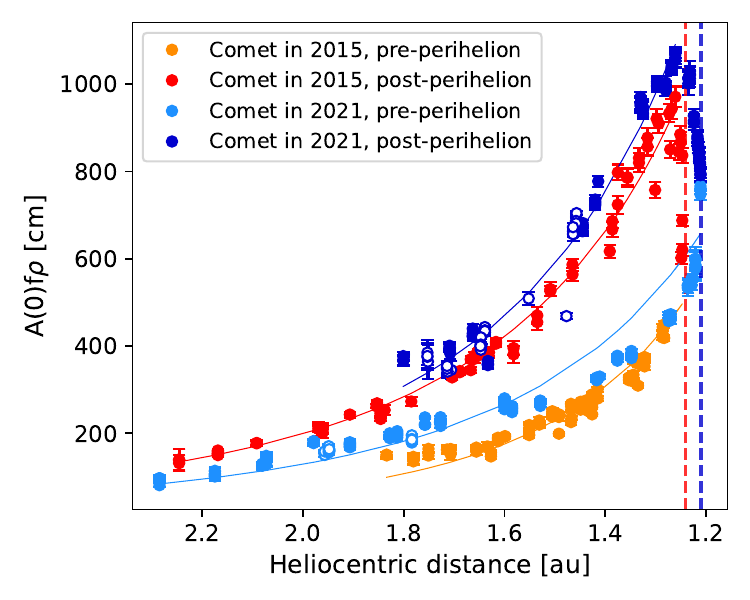} \\
        \includegraphics[width=\linewidth]{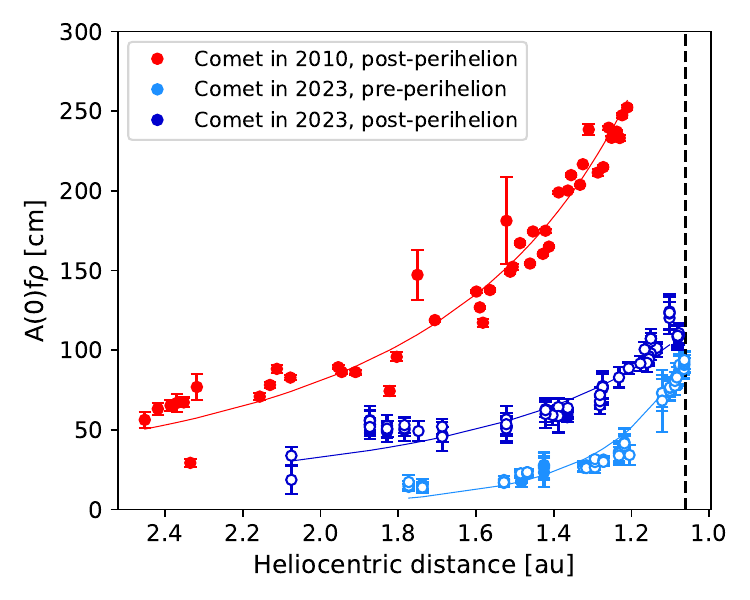}
    \caption{Evolution of the $A(0)f\rho$ parameter of comets 67P (up) and 103P (bottom) computed with the broadband R filter as a function of heliocentric distance during two perihelion passages. The dotted lines indicate the perihelion. The colored curves show the power-law fit applied on datasets. Filled and open points stand for TS and TN data, respectively.}
    \label{fig:afrhoR_67P_and_103P}
\end{figure}

To compare the dust activity of the comets from one passage to the other, we used the $A(0)f\rho$ values computed with the broadband R filter at each passage and plotted them together (see Fig. \ref{fig:afrhoR_67P_and_103P}). For both passages, we fit separately pre-maximum and post-maximum data with a power-law $A(0)f\rho \propto r_h^n$ to retrieve the exponent $n$ (also called the activity index), which gives the $r_h$-dependence of the dust activity. The results of the fit are shown on the plots, and the derived power-law exponents are given in Table \ref{tab:fit_afrho}. For both comets, our $n$ values are consistent with those generally obtained for Jupiter-family comets (see, for example, the survey of \cite{2024PSJ.....5...25G} who found average activity indices $n=-6.2\pm5.5$ before perihelion and $n=-3.6\pm3.4$ after perihelion). For 67P, we observe a slightly higher dust activity in 2021 than in 2015. The post-maximum activity indices are practically the same in both passages, while the pre-maximum index measured in 2015 is slightly higher than in 2021, indicating a faster increase in the activity during the pre-perihelion phase. Regarding the dust activity of 103P, it is clearly higher during the first passage than during the second one, at least in the post-perihelion phase of the comet. The post-maximum $r_h$-dependence is also slightly higher in 2010 than in 2023. 

\begin{table}
\centering
\caption{Activity indices $n$ derived from the power-law $A(0)f\rho \propto r_h^n$ adjusted on the $A(0)f\rho$ curves of comets 67P and 103P obtained with the R filter.}  
    \begin{tabular}{l >{\centering\arraybackslash}m{3cm} >{\centering\arraybackslash}m{3cm}}
    \hline
        & \multicolumn{2}{c}{67P} \\ 
        & \small{Before activity peak} & \small{After activity peak} \\
    \hline
    \hline
        2015 & $-4.18 \pm 0.03$ & $-3.37 \pm 0.02$ \\
        2021 & $-3.24 \pm 0.01$ & $-3.55 \pm 0.02$ \\
    \hline
        & \multicolumn{2}{c}{103P} \\ 
        & \small{Before activity peak} & \small{After activity peak} \\
    \hline
    \hline
        2010 & $-$ & $-2.30 \pm 0.01$ \\
        2023 & $-5.03 \pm 0.08$ & $-1.93 \pm 0.06$ \\
    \hline
    \end{tabular}  
\label{tab:fit_afrho}
\end{table}

\begin{figure}[htbp]
    \centering
        \includegraphics[width=\linewidth]{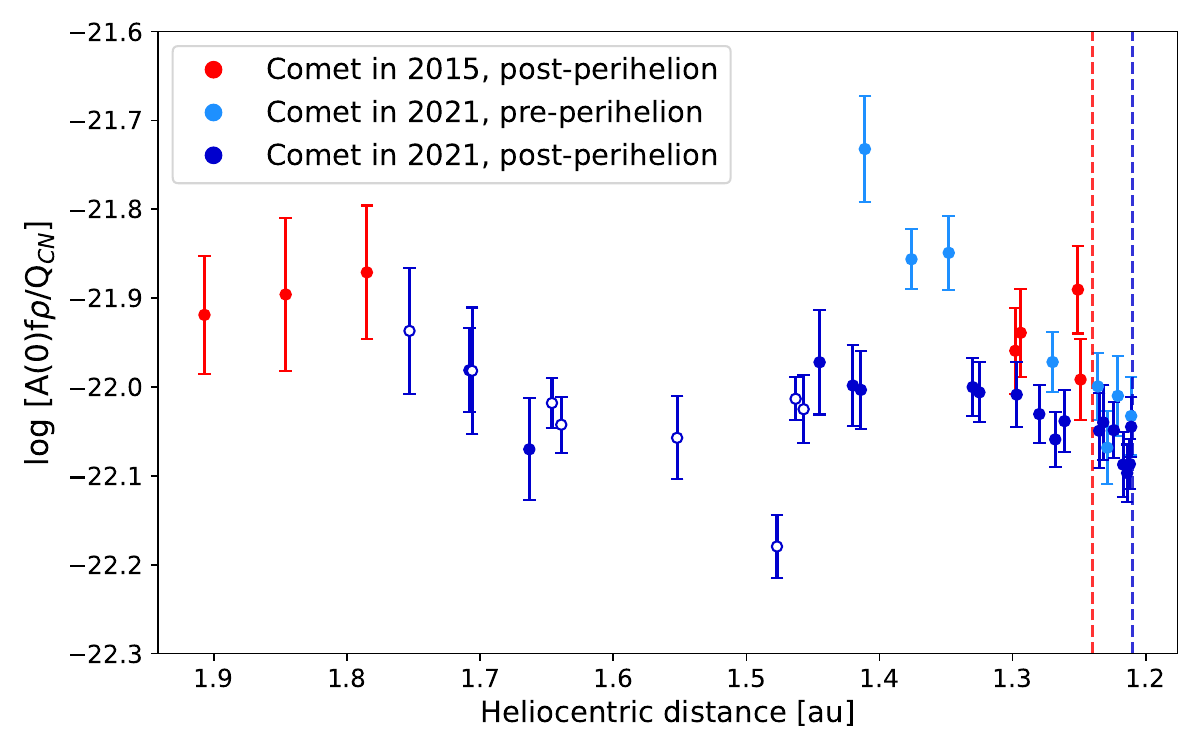} \\
        \includegraphics[width=\linewidth]{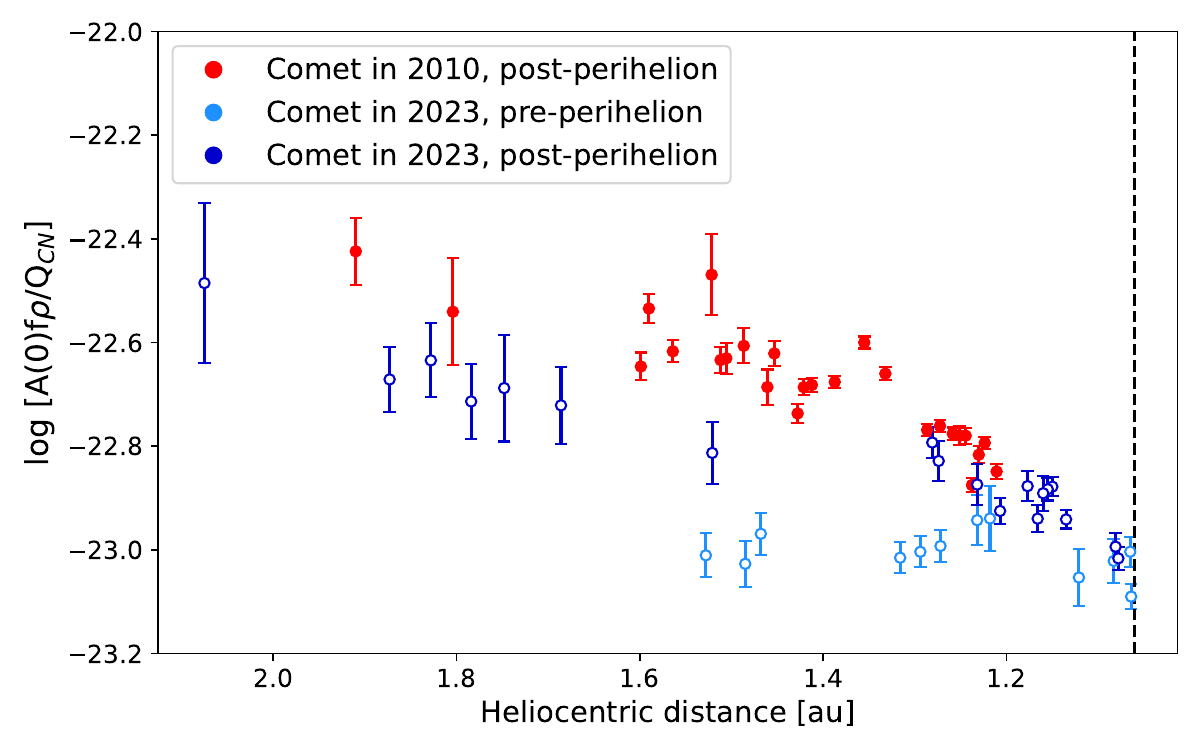}
    \caption{Evolution of the dust-to-gas ratios of comets 67P (up) and 103P (bottom) as a function of the heliocentric distance. The $A(0)f\rho$ values used here were computed with the broadband R filter.}
    \label{fig:log_afrho_QCN}
\end{figure}

Finally, we computed dust-to-gas ratios of both comets to get more information about their composition. In particular, this is useful to know whether a comet is rather dust-rich or gas-rich. For these ratios, we used the CN production rates as well as the $A(0)f\rho$ values obtained with the R filter. The results are shown in Fig. \ref{fig:log_afrho_QCN}. In both cases, we observe a small decrease in the ratios during the pre-perihelion phase and an increase after perihelion, although this variation is not clear for 67P. These comets would thus become slightly dustier as they approach the Sun. We also notice that 67P is a relatively dust-rich comet as compared to 103P. Moreover, the ratios of both comets seem to be in good agreement from one passage to the other, indicating their surface properties did not change between perihelia.

\section{Discussion}
\label{section:discu}

A probable explanation for the slight increase in cometary activity (gas production and dust outgassing) observed for 67P during its second perihelion passage as compared to the first one could simply be the small change in the comet orbit between the two passages. Indeed, the heliocentric distance at perihelion was of 1.24 au in 2015 and of 1.21 au in 2021. Since the comet was a little closer to the Sun during the second apparition, the more intense solar heat may have caused slightly greater gas sublimation. Moreover, 67P is an old comet that has already lost a lot of material during its successive passages near the Sun. A very small part of its nucleus still contains volatile elements and can therefore become active when it approaches the sun (see, for example, \cite{2019A&A...630A..18A} who modeled the active fraction of 67P's nucleus mainly based on data collected by Rosetta instruments and found a maximum value of around 10\% in two southern regions of the comet, rising up to 25-35\% at perihelion). Therefore, unless the comet's orbit changes (as in this case), we do not expect to see any change in its activity. The behavior of 67P is actually quite similar to the Jupiter-family comet 45P/Honda–Mrkos–Pajdusakova. Due to its perihelion distance being very small (0.53 au), large amounts of gas and dust are expected to be produced in the coma when the comet passes near the Sun, leading to an important mass loss and thus an important decrease in activity from one passage to another. However, \cite{2020PSJ.....1...72C} have only observed a small decrease in the activity of the comet over four perihelion passages (from 2001 to 2017). It is the opposite for 103P, which is a young hyperactive (i.e., its level of activity is high compared to what is expected based on the nucleus size) comet \citep{2011Sci...332.1396A}. For instance, \cite{2004A&A...419..375G} have estimated that the active fraction of 103P's nucleus reached 100\% at perihelion during its 1991 passage. For this comet, we observed significantly lower gas and dust activities during the second passage than during the first one. This could be due to an important loss of materials undergone by the comet between 2010 and 2023, which is expected since 103P has a surprisingly high level of activity. Moreover, we emphasize that the comet passed close to the Sun in 2017, although we were not able to observe this passage, and this also caused the comet to lose volatile iced molecules and dust particles. Another possible explanation for the observed decrease in 103P activity would be a change in the orientation of the comet with respect to the Sun. Indeed, some regions of the comet surface can contain more or less volatile elements and/or dust grains than others. Thus, according to the comet region that faces the Sun at perihelion, the level of activity can differ. However, in their survey on the evolution of water production activity of three periodic comets, \cite{2020PSJ.....1...72C} gathered the water production rates of comet 103P from various sources, computed for the 1991, 1997, and 2010 perihelion passages. Although these production rates have been obtained through different methods, a decreasing trend is unambiguously detectable at each passage. The authors also highlighted the same behavior for comet 46P/Wirtanen, another young Jupiter-family comet, for its apparitions in 1997, 2002, 2008, and 2018. Since the cometary activity of 103P decreased at each passage near the Sun from 1991 to the present day (without taking into account the 2004 and 2017 passages when the comet could not be observed because of poor geometry), we can consider that our first hypothesis is the most likely. 

An important part of the study of comets consists of determining whether their production rate ratios are constant or evolving with time. Constant ratios could indeed indicate that they are pristine. If so, then comets might have kept the same composition since the formation of the Solar System, and they could be used as clues to probe the chemical composition of the early Solar System. In this work, we classify both 67P and 103P as typical comets in terms of abundance in carbonated species and find no difference in the C$_2$-to-CN ratios from one passage to another. For 103P, this is rather consistent with the result obtained by \cite{1995Icar..118..223A} who found $\log[Q_{C_2}/Q_{CN}] = 0.08 \pm 0.02$ at $r_h = 1.04$ au for the 1991 passage of the comet and thus classified it as typical. However, concerning 67P, its classification as a typical comet has not always been supported. Indeed, \cite{1995Icar..118..223A} had found that it was carbon-depleted ($\log[Q_{C_2}/Q_{CN}] = -0.31 \pm 0.03$ at $r_h = 1.38$ au) based on observations they performed during the 1982 perihelion passage. Other authors have also reached this conclusion. We can cite, for example, \cite{2006Icar..181..442S} for the 1982 and 1995 passages ($\log[Q_{C_2}/Q_{CN}] = -0.21 \pm 0.27$ at $r_h = 1.35$ au), and \cite{2004A&A...422L..19S} for the 2002 passage ($\log[Q_{C_2}/Q_{CN}] < -0.21$ at $r_h = 2.9$ au). These authors have classified 67P as slightly depleted according to A'Hearn's criterion. However, more recent analyses actually revealed different results, which were more consistent with ours. See for instance \cite{1992Icar...98..151C} for the 1982 passage of the comet, \cite{2011A&A...525A..36L} (mean $\log[Q_{C_2}/Q_{CN}] = -0.10$ for $r_h=1.25$ au) for the 2009 passage, and \cite{2017MNRAS.469S.222O} (mean $\log[Q_{C_2}/Q_{CN}] = -0.08$ for $r_h$ between 1.25 and 1.30 au) for the 2015 passage. Both found a typical chemical composition. It is actually complicated to compare such measurements between various studies, due to the differences in the observation and computation methods used by each author. For example, \cite{1992Icar...98..151C} used different parameters for the Haser model, and \cite{2004A&A...422L..19S} computed gas production rates using a vectorial model instead of the Haser model. Some authors have also used non-homogeneous datasets, with images collected with different telescopes or filters, for their survey. This can explain the disparity in the results. In this work, we were able to accurately compare our measurements because we only used data collected from one single telescope, and for both passages of the two comets, the computation methods and model parameters remained unchanged, leading to highly homogeneous datasets. These results have also been compared with spectroscopic measurements, and they are consistent. Neither 67P nor 103P show any changes in their C$_2$-to-CN ratios from one passage to the other, although their cometary activity evolves. This indicates that there has been no change (or any change we could measure) in the chemical composition of the comets between their two apparitions. Moreover, this result is reinforced by the fact that we also do not observe any change in the dust-to-gas ratios of the comets from one passage to the other.


\section{Summary and conclusion}

The light curves of both comets reveal a steady evolution of their activity, characterized by a decreasing magnitude during the pre-perihelion phase and an increasing magnitude during the post-perihelion phase, with a minimum reached a few days after perihelion as generally expected for comets. For the whole monitoring of both 67P and 103P, we did not observe any noteworthy outbursts. These Jupiter-family comets exhibit steeper slopes than those of long-period and dynamically new comets, suggesting distinct activity behaviors. It is therefore essential that the number of such measurements using high-quality light curves be increased to confirm the statistical robustness of this trend, an objective we will pursue in particular through observations with TRAPPIST.

The analysis of coma dust colors showed almost constant color indices over a large heliocentric distance range, except for the B$-$R index, which slightly varies during the second passage of 103P. This variation can be attributed to the gas contamination in the V broadband filter. For both comets, the mean color indices computed at each passage are consistent with the mean values obtained for active Jupiter-family comets, and we found no significant difference in these mean values from one passage to the other, which indicates that the cometary dust properties remained unchanged. 

For 67P, we measured a slightly higher gas and dust activity during the second apparition, as compared to the first one. Since 67P is a rather old comet that is not really expected to evolve, this increase in its activity is probably due to the small decrease (0.3 au) in the perihelion distance that was observed between 2015 and 2021, and that caused the heating of the comet to be more important. For 103P, we measured a clearly lower activity during the second passage for the same perihelion distance, which is due to the high activity level of this young comet, leading to an important loss of materials at each passage, and so to a considerable drop in activity. For both comets, our measurements of the gas production rates are in very good agreement with the spectroscopic measurements performed with the HCT, confirming our values and methods. 

Following the criterion defined by \cite{1995Icar..118..223A}, we found a typical composition for both comets. Moreover, we did not observe any significant change in these ratios from one passage to the other, which indicates that the chemical composition of comets might remain preserved even if they undergo several passages near the Sun. However, this result is still controversial in the literature, and a more extended study, based on monitoring of periodic comets over more than two perihelion passages with the same observation and computation methods, should be conducted to be able to draw a conclusion about this. This is unfortunately not an easy task, as many years separate different perihelia, and the comets are not always well visible for each passage.

Finally, we measured that 67P is a relatively dust-rich comet as compared to 103P. For both comets, we found no change in the dust-to-gas ratios measured from one apparition to the other, which reinforces our hypothesis about the constant composition of comets over time.


\section*{Data availability}
Tables containing the gas production rates and $Af\rho$ parameter of comets 67P and 103P computed from TRAPPIST data are only available in electronic form at the CDS via anonymous ftp to cdsarc.u-strasbg.fr (130.79.128.5) or via \url{http://cdsweb.u-strasbg.fr/cgi-bin/qcat?J/A+A/}.

\begin{acknowledgements}
This publication uses data products from the TRAPPIST project, under the scientific direction of Emmanuel Jehin, Director of Research at the Belgian National Fund for Scientific Research (F.R.S.-FNRS). TRAPPIST-South is funded by F.R.S.-FNRS under grant PDR T.0120.21, and TRAPPIST-North is funded by the University of Liège in collaboration with Cadi Ayyad University of Marrakech. M. Vander Donckt acknowledges support from the French-speaking Community of Belgium through its FRIA grant. K. Aravind acknowledges support from the Wallonia-Brussels International (WBI) grant. This work is a result of the bilateral Belgo-Indian projects on Precision Astronomical Spectroscopy for Stellar and Solar System bodies, BIPASS, funded by the Belgian Federal Science Policy Office (BELSPO, Govt. of Belgium; BL/33/IN22\_BIPASS) and the International Division, Department of Science and Technology (DST, Govt. of India; DST/INT/BELG/P-01/2021 (G)). S. Hmiddouch acknowledges funding from the Belgian Academy for Research and Higher Education (ARES). The authors thank NASA, David Schleicher, and the Lowell Observatory for the loan of a set of HB comet filters.
\end{acknowledgements}

\bibliographystyle{aa}
\bibliography{references}


\end{document}